\documentclass[fleqn,10pt]{wlscirep} 
\usepackage[utf8]{inputenc}
\usepackage[T1]{fontenc}
\title{Distinguishing mechanisms of social contagion from local network view}

\usepackage{bm}
\usepackage{color, soul, colortbl} 

\usepackage{xfrac}
\usepackage[overload]{empheq}
\usepackage{mathtools}
\usepackage{cases}
\usepackage{csvsimple}

\usepackage[switch]{lineno} 
\usepackage{bbold}
\usepackage{subcaption}
\usepackage{xcolor}

\definecolor{brickred}{rgb}{0.8, 0.25, 0.33}

\author[1]{Elsa Andres}
\author[1,2]{Gergely Ódor}
\author[3,4]{Iacopo Iacopini}
\author[1,2,*]{Márton Karsai}
\affil[1]{Department of Network and Data Science, Central European University, Vienna, 1100, Austria}
\affil[2]{National Laboratory for Health Security, HUN-REN Alfréd Rényi Institute of Mathematics, Budapest, 1053, Hungary}
\affil[3]{Network Science Institute, Northeastern University London, London, E1W 1LP, United Kingdom}
\affil[4]{Department of Physics, Northeastern University, Boston, MA 02115, USA}

\affil[*]{karsaim@ceu.edu}

\begin{abstract}

The adoption of individual behavioural patterns is largely determined by stimuli arriving from peers via social interactions or from external sources. Based on these influences, individuals are commonly assumed to follow simple or complex adoption rules, inducing social contagion processes. In reality, multiple adoption rules may coexist even within the same social contagion process, introducing additional complexity into the spreading phenomena. Our goal is to understand whether coexisting adoption mechanisms can be distinguished from a microscopic view, at the egocentric network level, without requiring global information about the underlying network, or the unfolding spreading process. We formulate this question as a classification problem, and study it through a Bayesian likelihood approach and with random forest classifiers in various synthetic and data-driven experiments. This study offers a novel perspective on the observations of propagation processes at the egocentric level and a better understanding of landmark contagion mechanisms from a local view.

\end{abstract}

\begin{document}

\flushbottom
\maketitle
%
%



\section*{Introduction}

We influence our peers by our conduct and interactions, thereby impacting their decisions to follow behavioural patterns similar to ours. Such patterns, mediated by social influence, may propagate as a spreading process and lead to macroscopic phenomena of mass adoption of products, ideas, beliefs, or information cascades~\cite{castellano2009statistical, barrat2008dynamical, jose2022covid, christakis2008collective}. The relevance of social spreading phenomena has been previously identified \cite{bass1969new, rogers2014diffusion}, and arguably explained by simple decision mechanisms on well-mixed populations \cite{maki1973mathematical, daley1965stochastic, granovetter1973strength, schelling1971dynamic}. Meanwhile, the importance of social networks has also been recognised~\cite{chen2022information, centola2010spread, moreno2004dynamics}, as they effectively encode the underlying structure along which social influence travels. Their structure could critically influence the global outcome of social spreading phenomena unfolding on top of them~\cite{barrat2008dynamical,watts2002simple}. This finding is especially true for temporal networks~\cite{holme2012temporal}, which capture both the structure and the time of interactions between connected peers, whose time varying links represent possible events of direct social influence~\cite{masuda2017introduction,unicomb2021dynamics}.\\

Models of social contagion commonly describe the spreading dynamics as a binary state process~\cite{gleeson2013binary}, in which individuals are identified as nodes of a social network that can be in different states; susceptible nodes (also called ignorants) may adopt a given behaviour and become ``infected'' ---borrowing the term from the literature of infectious disease modelling---, or in other words spreaders, or adopters\footnote{Note that in this manuscript, we would use these terms interchangeably} through a cognitive process driven by a variety of contagion mechanisms. One family of mechanisms \cite{maki1973mathematical, daley1965stochastic, castellano2009statistical}, commonly termed \emph{simple contagion} in the social science literature \cite{gleeson2016effects, crane2008robust, sreenivasan2016information, gleeson2014competition, centola2007complex}, resembles biological epidemic processes; each interaction between a susceptible node and an infectious one may independently result in an infection event with a predetermined probability, leading to gradually evolving global adoption curves ~\cite{pastor2015epidemic}.

There is however plenty of empirical evidence suggesting, in certain contexts, an alternative mechanism, called \emph{complex contagion}~\cite{centola2007complex}. In this case, exposures are not independent, but peer pressure can impact in a non-linear way the individual infection probability, for example by accumulating influence towards an individual adoption threshold~\cite{granovetter1978threshold, schelling1971dynamic, watts2002simple}.
Depending on the model parameters, the complex contagion mechanism may lead to a cascading phenomenon, where mass infection emerges over a short period of time. This was first shown on networks by Watts~\cite{watts2002simple}, while several follow-up studies explored a rich family of similar phenomena in multi-layer \cite{de2018fundamentals, min2023threshold, karsai2016local}, weighted \cite{unicomb2018threshold, li2019universal} or temporal networks \cite{karimi2013threshold, unicomb2021dynamics}, demonstrating their relevance in real-world settings~\cite{hodas2014simple, vasconcelos2019consensus, monsted2017evidence, karsai2014complex, borge2011structural}. In this manuscript, we will use the threshold model~\cite{watts2002simple} as a paradigmatic mechanism of complex contagion.


Simple and complex contagion capture network-based adoptions, however, social influence may not always spread on an observable network (e.g., advertisements, news or policy recommendations, etc.). We take such external influences into account by also considering a third mechanism, called \emph{spontaneous adoption} \cite{karsai2014complex, hill2010infectious, toole2012modeling}. Although spontaneous adoption is agnostic to the underlying network structure, infection patterns via the other two mechanisms depend non-trivially on several network and dynamical characters of an ego and its peers~\cite{ugander2012structural}. It has been shown that while simple contagion spreads easier on dense and degree-heterogeneous structures, with high-degree nodes early infected~\cite{yang2015large}, these properties mitigate complex contagion as the threshold of high-degree nodes can be hardly reached~\cite{barrat2008dynamical,cencetti2023distinguishing}. Moreover, while weak ties connecting densely connected communities act as facilitating bridges for simple contagion~\cite{granovetter1973strength}, they slow down complex contagion cascades~\cite{centola2007complex}, as they likely deliver non-reinforced social influence to susceptible individuals. In addition, the timing and the order of infection stimuli, their concurrency, and the bursty dynamics of interactions~\cite{karsai2018bursty, masuda2021concurrency, karimi2013threshold} between individuals and their peers influence the adoption dynamics and the macroscopic dynamical outcome of the spreading process as a whole.\\

Acting alone, all social contagion mechanisms may lead to differentiable infection dynamics at the global scale. In this direction, while distinguishing mechanisms solely from the overall infection dynamics remains a challenge~\cite{hebert2020macroscopic}, recent methods combining spreading dynamics and network information~\cite{cencetti2023distinguishing}, or considering the timing of peer stimuli~\cite{monsted2017evidence} led to promising results. However, these studies commonly make two assumptions limiting their applicability in real-world scenarios. First, they expect full knowledge about both the underlying network structure and the spreading dynamics. Indeed, this is a strong assumption in common real-world scenarios, where information about infection events is typically incomplete or limited to local knowledge, possibly obtained only about an adopting ego and its peers. Second, these studies assume that all individuals follow the same single adoption mechanism; either simple or complex contagion. In contrast, it has been argued that the mechanism driving one's decision to adopt a behaviour during an unfolding social contagion may depend on the intrinsic susceptibility of an individual to the actual behavioural form and the properties of the propagation process itself \cite{gladwell2006tipping, aral2017exercise}. Thus, each single adoption event may be driven by different mechanisms that jointly depend on personal factors~\cite{state2015diffusion, karampourniotis2015impact} (heterogeneous susceptibility and predisposition), the properties of the item being adopted (Gladwell's stickiness~\cite{gladwell2006tipping}), and the particular context (environment, time of adoption, other external factors).\\ 

In this study, we distinguish between simple, complex and spontaneous contagion mechanisms by addressing both the issue of limited data availability and the challenge that a single social contagion process may involve multiple adoption mechanisms \cite{min2018competing}. We frame this question as a classification problem and explore solutions based on Bayesian likelihood and random forest approaches. These methods are developed and tested on extensive synthetic simulations, encompassing different spreading scenarios and underlying social structures, ranging from fully controlled experiments to empirical spreading cases observed on Twitter (currently called X). Our ultimate goal is to uncover the fundamental limits of distinguishability of these mechanisms, and to propose solutions that can be readily used in real-world scenarios aimed at understanding social contagion phenomena.


\section*{Results}

\subsection*{Different mechanisms of social contagion}

\begin{figure*}[h!]
\centering
\includegraphics[width=\textwidth, trim={0cm 6cm 0cm 0cm}]{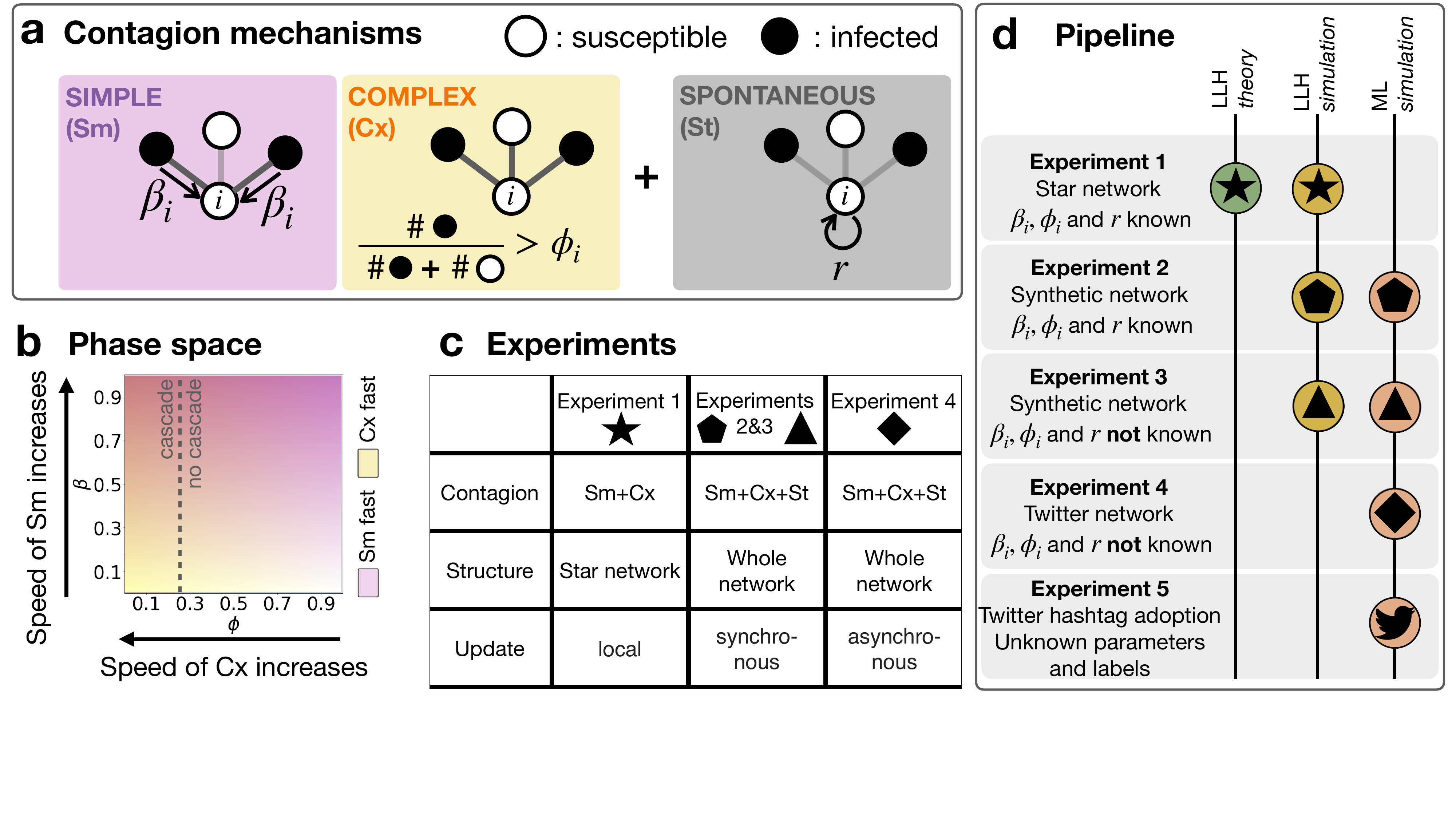}
\caption{Overview of experimental setups. (a) Illustration of the three contagion mechanisms that are subject for inference: \emph{simple contagion} parameterised by the adoption probability $\beta_i$ through a single stimuli; \emph{complex contagion} parameterised by the threshold $\phi_i$ of necessary fraction of adopter neighbours to induce the adoption event; and \emph{spontaneous adoption} that occurs with probability $r$. (b) The parameter space $(\beta,\phi)$ and the speed dependence of the simple and complex contagion processes. (c) The different experimental setups that include the considered contagion mechanisms, the complexity of the underlying network, and model update rules. (d) Schematic pipeline for the application of the log-likelihood (LLH) and random forest machine learning (ML) classification approaches to the different experiments.}
\label{schema}
\end{figure*}

We study adoption processes on networks, where individuals are represented as a set of nodes $V$, and their interactions as a set of links $E$. The number of nodes connecting to a node $i$ (i.e., the number of neighbors) is called the degree of $i$, denoted by $k_i$. The most common way to model propagation dynamics on a networked population is to assign a state to every node, which characterises its status with respect to the propagation~\cite{gleeson2013binary}; a node is either susceptible (S), meaning that it has not yet been reached by the contagion process, or infected (I), if it has already been reached, and thus it can infect others.

We consider three infection mechanisms that can change the state of a node from susceptible to infected (cf Figure~\ref{schema}a).
As for the \emph{simple contagion} (Sm) mechanism, we build on the Susceptible-Infected (SI) model, introduced first in epidemiology \cite{stattner2011social} and later to characterise the adoption of social behaviours \cite{gleeson2016effects, crane2008robust, sreenivasan2016information, gleeson2014competition}. In this model a susceptible node can become independently infected with a fixed probability during each interaction with an infectious neighbour. Here, we assume that
at every time step a susceptible node $i$ could acquire an infection from each infectious node in its neighbourhood with its node-dependent probability $\beta_i\in[0, 1]$ (that could thus be considered alike a heterogeneous susceptibility). After a gradual contamination of the network, the macroscopic steady state of an SI contagion process is reached when all nodes become infected.
The \emph{Complex contagion} (Cx) mechanism breaks the linearity of the contagion by introducing social reinforcement effects, often found in behavioural patterns: it is the \textit{combined} influence arriving from the neighbours of an ego node, which triggers the adoption. Here, we consider this mechanism by employing a conventional deterministic threshold model introduced by Watts~\cite{watts2002simple}, where each susceptible node $i$ becomes infected as soon as its fraction of infectious neighbours exceeds a preassigned intrinsic threshold $\phi_i\in[0, 1]$. This threshold model is known to exhibit rapid cascading behavior if the necessary conditions on the average degree and the infection threshold are met \cite{watts2002simple}.
%
%
Parameters $\beta_i$ and $\phi_i$ are crucial in shaping the propagation dynamics. High values of $\beta_i$ lead to faster adoption via Sm, while low values of $\phi_i$ accelerate the adoption rate via Cx, as individual thresholds become easier to reach (see Figure \ref{schema}b and also Supplementary Material 1).
Finally, we implement a third adoption mechanism called \emph{spontaneous adoption} (St), which models external effects; every susceptible node becomes infected with probability $r$ during any time steps of the process, independently of the state of its neighbours.

The backbone of the paper is a series of four experiments (Figure~\ref{schema}c), where we tackle the problem of distinguishing simulated Sm, Cx and St processes based on the infection times of an ego node and its neighbours. The experiments cover a wide range of scenarios, from the simplest configuration on disjoint star networks with $\beta_i$ and $\phi_i$ known to the estimator (Experiment 1), to the most involved setup, simulated with co-existing, asynchronous update mechanisms with unknown parameters (Experiment 4). In each experiment, we distinguish the adoption processes using a Bayesian maximum likelihood approach and a random forest classifier, whenever the method is applicable (Figure~\ref{schema}d). The Bayesian likelihood approach features theoretical guarantees, and the possibility to include prior knowledge about the underlying processes \cite{price2018bayesian}. However, likelihood-based approaches may not be robust if they cannot capture precisely the data from the assumed generative process \cite{hansen2001model}. In contrast, random forest classifiers tend to be more robust even if the dataset does not fit perfectly to the model, while falling short on the interpretability of the results. Finally, after highlighting the strengths and weaknesses of the two classification approaches, we apply the random forest classifier to real ego-level datasets collected from the Twitter (now called X) micro-blogging and social networking platform.

\subsection*{Process classification with known parameters}
We start approaching the proposed classification task in the most elementary case, that is when the parameters $\{\beta_i\}_{i \in N}$, $\{\phi_i\}_{i \in N}$ and $r$ governing the spreading processes are known to the classifier. Even though such information is not available in practical real-world scenarios, this setup represents an ideal starting point to understand the performance of the classifiers in a simple and controlled synthetic context.

\subsubsection*{Contagion on egocentric networks}

\begin{figure*}[h!]
\centering
\includegraphics[width=180mm]{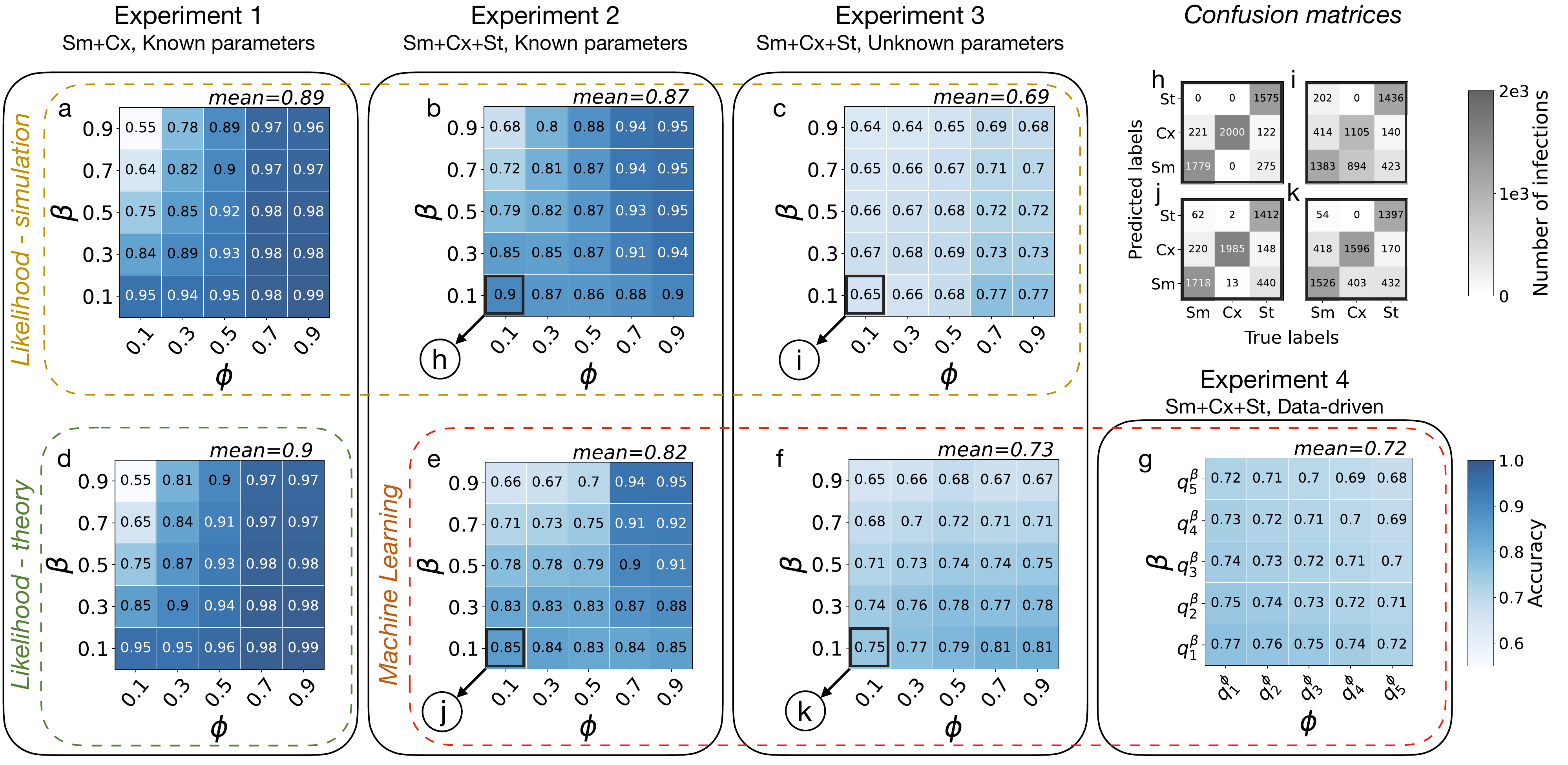}
\caption{Classification accuracy values of the likelihood method (green rectangle (d) when it is obtained theoretically and yellow rectangle (a-c) when it is obtain by simulation) and of the random forest method (red rectangle (e-g)). Results in the same column are obtained on the same Experiment produced by synthetic models, with model complexity increasing from left to right. In panel (g), the notation $q^{parameter}_n$ represents the $n^{th}$ quintile of the parameter distribution. Panels (h-k) show the confusion matrices associated to the highlighted pairs ($\beta$, $\phi$) from Experiments 2 and 3. In general, classification accuracy decreases with increasing model complexity, but the accuracy remains well-above the random baseline (0.5 for Experiment 1 and 0.33 for Experiments 2-4). Within one experiment-method pair, accuracy increases with $\phi$ and decreases with $\beta$, which agrees with our intuition that the Sm and the Cx are most difficult to distinguish when both contagions propagate fast in the network.}
\label{accuracies_methods}
\end{figure*}

\paragraph{Experiment 1.} As we aim at classifying contagion mechanisms relying solely on the information available at the level of an ego node and its neighbours, the simplest setting to consider is the case of contagion processes that spread on disjoint star structures that are not part of a larger network structure. To isolate the mechanism of the ego node only, we assume that all the neighbours undergo a spontaneous adoption (St mechanism), while the ego can adopt via simple or complex mechanisms, which are randomly assigned at the beginning of each simulation, as well as the $\beta_i$ and $\phi_i$ parameters controlling the contagion of each ego node. 

After simulating the contagion process for $T$ timesteps, we feed the classification algorithm with the trajectory $\{\sigma_i(t)\}_{t=0}^T$ that takes values 0 (S) or 1 (I) and tracks the status of each ego node $i$ at each timestep $t$. In order to assess whether the trajectory of an ego has been generated by the Sm or Cx mechanism, we formulate the classification problem under a Bayesian likelihood framework.
Since both contagion processes are Markovian (i.e. the state of the system at a given time only depends on the previous timestep), we can write, for each node $i$, the likelihood for an observed process to be generated by each mechanism $\mathcal{X} \in \{Sm, Cx\}$ with parameters $\{\beta, \phi\}$ as the product of the probabilities:
\begin{equation}
  \mathcal{L}_i(\mathcal{X}) = \prod_{t=0}^{T} \mathbf{P}(\sigma_{i}(t+1)|\sigma_{i, nb}(t), \mathcal{X}, \{\beta, \phi\})  ,
  \label{llh_tiago}
\end{equation}
where $\sigma_{i, nb}(t)$ denotes the trajectories of the ego node and of its neighbours. An observed adoption could then be attributed to the mechanism having the highest likelihood (more details are given in “Likelihood calculations” of the Methods section).

Assuming that the star networks have degrees $k$ drawn from a binomial distribution, we display in the heatmap of Figure \ref{accuracies_methods}a the obtained accuracies (proportion of well-classified nodes) as a function of the respective pair of parameters $(\beta, \phi)$ that generated the simulations. We obtain relatively high accuracy values ---with a mean of 0.9--- over the whole parameter space, with the exception of the portion of the space where Sm and Cx both evolve fast, which corresponds to the parameter extreme when $\beta\rightarrow 1$ and $\phi\rightarrow 0$. In this case, Sm and Cx are very difficult to distinguish; in both cases, the ego node becomes infected most likely one timestep after its first neighbour adopts. This parameter range also corresponds to the least distinguishable scenario at the level of the global epidemic curves, as they both evolve rapidly even in populations with homogeneous adoption mechanisms (Supplementary Material Figure S1).
In this range, the lowest classification accuracy is around $0.55$, which is still slightly above the expected accuracy of a random classifier $0.5$. Notably, the two processes are highly distinguishable in the opposite case, when $\beta=0.1$ and $\phi=0.9$. In this other extreme, $\phi$ is so high that Cx adoptions are possible only once most of the neighbours of the adopting ego have been spontaneously infected. At the same time, Sm adoptions are still possible via repeated stimuli from a few neighbours, making the two processes easier to distinguish.

A major advantage of this stylised setup on disjoint degree-$k$ star networks is that the Bayesian classification accuracy can be approximated analytically as
\begin{equation}
\label{eq:analytic_result}
    \begin{aligned}
    ACC(k,\beta,\phi,r) 
    & \approx 1- \frac12 \left(\prod_{n=1}^{\lfloor k\phi \rfloor} \frac{p_n-p_nb_n}{b_n+p_n-p_nb_n} \right) b_{\lceil k\phi \rceil},
    \end{aligned}
\end{equation}
with $p_n=1-(1-r)^{k-n}$ and $b_n=1-(1-\beta)^{n}$ (see Methods for the details of the calculation).
Comparing the theoretically estimated accuracies from Eq.~\eqref{eq:analytic_result} (visualised in Figure \ref{accuracies_methods}d) with the simulation outcomes (Figure \ref{accuracies_methods}a), we observe a very close match, with a maximum difference of 0.01.

Overall, Experiment 1 features a high classification accuracy and precise analytical results, while making strong assumptions on the network structure and the adoption mechanisms. Since the Bayesian likelihood approach matches the underlying model exactly, it is an optimal estimator, and we omit the application of the random forest approach in this setup. However, since this setting also neglects some of the most important features of realistic social contagions and social structures, it can only be considered as the simplest solvable reference model to be compared with more complex scenarios.



\subsubsection*{Contagion on random networks}

\paragraph{Experiment 2.} To generate a more realistic setting, we consider contagion mechanisms that spread over larger network structures. Most of the results in this section were obtained on the giant component of Erdős-Rényi random networks \cite{erdHos1960evolution} with 1000 nodes and an average degree of 4, but we also present results on random networks with degree heterogeneity, triadic closure and community structure with the same parameters. Similarly to Experiment 1, we randomly predetermine the contagion mechanism (simple or complex) for each node. This time, however, we allow each node to spontaneously adopt during the contagion process, regardless of their predetermined mechanism. This way the contagion does not vanish even on large networks with extreme Sm and Cx contagion parameters, but continues spreading following a linear dynamics. The modification also implies that, since nodes can adopt via the simple, the complex or the spontaneous mechanisms, our classification algorithms need to distinguish between the three hypotheses (see Methods).

In line with the approach of Experiment 1, we compute the likelihood that each adopter follows a specific contagion mechanism (see Eq.~\ref{llh_tiago}) based on the trajectories of the ego nodes and their neighbours. Since the assumption on the independent adoption of the neighbours of an ego does not hold anymore, the likelihood framework becomes an approximation (see “Likelihood calculations” of the Methods section for the detailed derivation). Nevertheless, accuracy values for the whole parameter space summarised in Figure \ref{accuracies_methods}b confirm that this approach can still perform well achieving a mean accuracy of 0.87 ---well above the expected accuracy of a random classifier ($0.33$).

Since the likelihood framework provides an approximate solution for Experiment 2, it calls for alternative approaches. After an extensive classification model selection (cf Supplementary Material 2), we selected a random forest approach as the consistently best performing classifier. In order to strike a balance between performance and interpretability, we train random forest classifiers on the same synthetic dataset as above. After testing on several structural and dynamical features of the ego and its neighbours, we identify eight relevant features for the classification that appear with distinct distributions for different infection mechanisms (cf Supplementary Material 3). These are (i) the degree, (ii) the proportion of infected neighbours, (iii) the number of infected neighbours, (iv) the sum of received stimuli, (v) the average number of received stimuli by neighbour, (vi) the standard deviation of per neighbour stimuli, (vii) the time since the first infected neighbour and (viii) the time since the last infected neighbour.

\begin{figure}[h!]
\centering
\includegraphics[width=.6\linewidth]{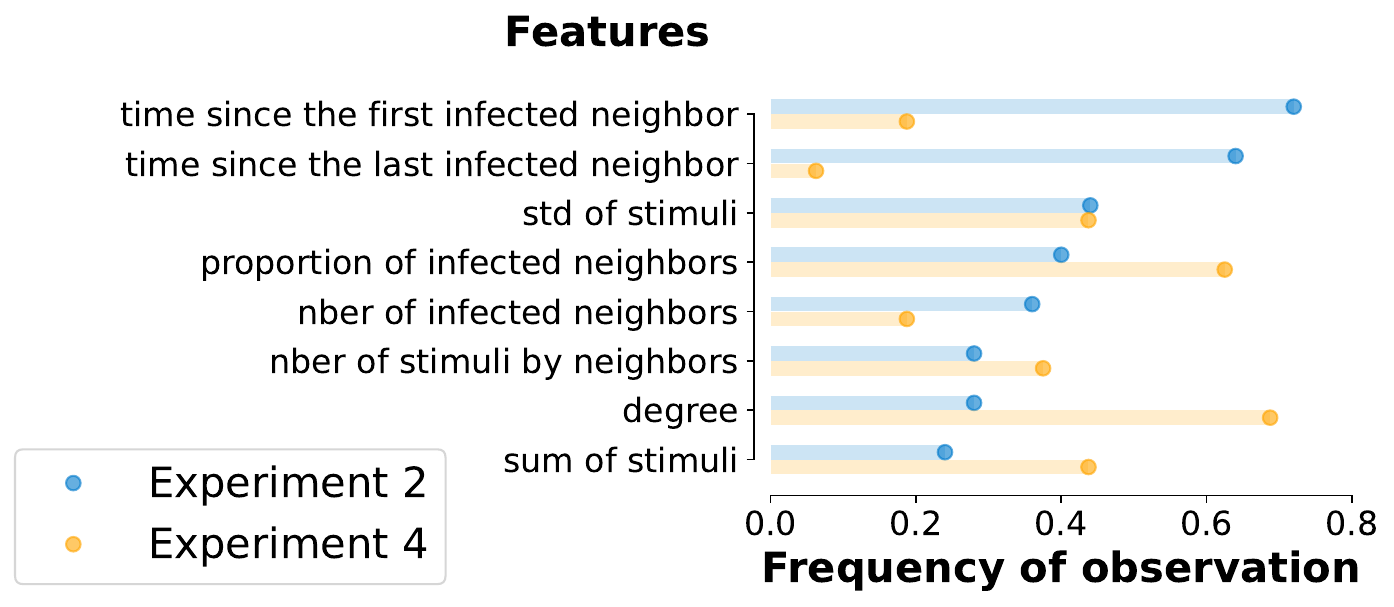}
\caption{Frequency of observation of the features used to train the random forest classifier among the top-3 most important ones across the full parameter space for Experiment 2 (blue) and Experiment 4 (orange). Frequencies are computed as the number of appearances normalised by the number of possible occurrences. The resulting most important features are the \emph{time since the first} and \emph{the last infected neighbour}.}
\label{best_features}
\end{figure}

We train a random forest model using these input features for each adopted node that appeared during a simulated contagion with Sm and Cx with parameters $\beta$ and $\phi$. The random forest approach provides very similar results (see Figure \ref{accuracies_methods}e) to the likelihood-based calculations (Figure \ref{accuracies_methods}b), only with slightly worse average accuracy $0.82$. According to the the confusion matrices shown in Figure \ref{accuracies_methods}h and j, while the two methods perform similarly in classifying simple contagion cases, the random forest misclassifies complex and spontaneous instances at a higher rate. Notably, given the interpretability of the trained random forest classifiers via feature importance, we can further restrict our original eight features to only three, and retain similar accuracies as before (see Supplementary Material 3). Interestingly, some feature subsets are consistently optimal across the full parameter space. This is reported in Figure \ref{best_features}, where we present the number of times a feature appears within the subset of the top-3 optimal features, normalised by the number of possible instances (parameter pairs $\beta, \phi$ in the phase space). Overall, the two most recurring features are the times since the first and the last infected neighbours. These can be also easily interpreted within the modelling framework: the time since the first infected neighbour cannot be too high for Sm, as that would mean too many repeated stimuli without an infection event, while for the threshold-based Cx the time since the last infected neighbour has to be necessarily one.

\subsection*{Process classification with unknown parameters}

Up to this point, all the investigated tasks assumed precise knowledge of the parameters $\beta$, $\phi$ and $r$ governing the different processes. However, in realistic scenarios, these need to be also inferred together with the contagion mechanisms, thus motivating the following experimental setup.

\paragraph{Experiment 3.} In this setting we classify the contagion instances from Experiment 2 assuming unknown contagion parameters, which means distinguishing mechanisms without knowledge on the parameters that governed them. In the Bayesian approach, we use the same equations to compute the likelihood that the contagion instance $i$ is simple, complex or spontaneous as before, except we also estimate the values of $\beta_i$, $\phi_i$ and $r$. We set the value of $\hat{\beta}_i$ as the inverse of the number of received stimuli by the node $i$, and the value of $\hat{\phi}_i$ as the proportion of infected neighbours at the time of the infection of node $i$. The value of $\hat{r}$ is calculated as the fraction of time spent by a node in the S state with at least one infected neighbour (see Methods for more details). 

Figure~\ref{accuracies_methods}c shows that we still classify the adoption mechanisms with high accuracy, especially considering the increased difficulty of the classification problem compared to the earlier settings. The mean accuracy was found to be $0.69$, well above the reference value of a random classifier (0.33). We observe the worse performance for low values of $\phi$, due to the high rate of confusion between complex and simple contagion cases (Figure \ref{accuracies_methods}i). Those nodes are generally infected just after the appearance of an infectious neighbour, making it difficult for the model to distinguish between the two peer-driven mechanisms Sm and Cx. The accuracy is the highest for large values of $\phi$ and low values of $\beta$. As before, we gain the most information about the processes when both of them are progressing slowly.

We also test the random forest approach in this experiment by using the same features used in Experiment 2, but training instead one unique model over the whole phase space ---as the parameters are not known anymore. Interestingly, this solution provides slightly more accurate results (see Figure \ref{accuracies_methods}f) than the likelihood method (see Figure \ref{accuracies_methods}c), especially for low values of $\beta$. Reading the confusion matrices (in Figure \ref{accuracies_methods}k and i resp.), this improvement mostly comes from the better classification of complex contagion instances, that were commonly classified as simple by the likelihood approach. Nevertheless, the overall accuracy of the random forest classifier is lower for Experiment 3 as compared to Experiment 2, which is expected, as the estimators receive less information.

Note that we conducted Experiment 2 and 3 on various types of random networks including Erd\H{o}s-R\'enyi~\cite{albert2002statistical} (presented above), Barab\'asi-Albert~\cite{barabasi1999emergence}, Watts-Strogatz~\cite{watts1998collective} and Stochastic Block Model~\cite{lee2019review} networks (see Supplementary Material 4) with very similar results. This suggests that the global network structure has limited impact on the local differentiation of contagion processes in each performed experiment.


\subsection*{Case study: adoption mechanisms on Twitter}

After demonstrating the validity of our methods in controlled synthetic settings, we now turn our focus towards real contagion processes to showcase the applicability of the devised approach to empirical scenarios. To this end, we rely on an ego-level dataset of adoptions from Twitter\cite{de2022measuring} (now called X), a micro-blogging and social networking platform, where users can follow each other, and share short messages, or tweets. The dataset contains all tweets posted by 8527 selected users (egos who are interested in French politics) and the people they follow (whom we call followees, or the members of the ego network) between May 1 2018 to May 31 2019 (for more details about the data collection see~\cite{de2020information}). This mounts up to a total of 1,844,978 timelines, i.e., the timely ordered personal stream of tweets posted by all these users. This dataset allows us to identify the time of adoption of a given hashtag by an ego together with the time of all incoming stimuli from its neighbours that previously posted the same hashtag. These tweets cover multiple topics, which may correspond to the spreading of various co-occurring social contagion processes. Since we are interested in analysing each contagion process separately, we filter messages that contain a given set of hashtags within the same topic.
We choose to focus on the hashtag \#GiletsJaunes and its variants\footnote{We target every user who has posted one of those hashtags: \#GiletsJaunes, \#giletsjaunes, \#Giletsjaunes, \#GiletJaune, \#Giletjaune, \#giletjaune, \#giletsjaune, \#Giletsjaune, \#GJ}, characterising a political uprising in France that induced a significant social contagion unfolding on Twitter. We first identify egos who adopted a related hashtag, and observe the posts of their followees over the preceding week, limiting in this way the effect of influence to the recent past only. As per the synthetic cases, we can define the degree of an ego as the number of its followees who have posted at least one tweet during the week preceding the adoption. In addition, user activity on Twitter is not linear in time ---as in our previous simulations--- but it is driven by circadian fluctuations, bursty patterns, and individual preferences. We thus move from real-time to event-time simulations. In this setting, a time step for an ego (leading to potential adoption cases) is counted as the number of tweets by the followees, regardless of weather they contain the hashtag of interest; every time an alter posts content containing the selected hashtag, the ego will receive a stimulus.



\begin{table}[t]
    \centering
    \begin{tabular}{|l|l|l|l|}
    \hline
        ~ & Sm & Cx & St \\ \hline
        Random forest & 970 & 349 & 4955 \\ \hline
        Likelihood & 4440 & 1447 & 387 \\ \hline
    \end{tabular}
    \caption{Number of instances of contagion mechanisms inferred by the likelihood and random forest methods on the \#GiletsJaunes Twitter dataset.}
    \label{table}
\end{table}

Empirical traces of social contagion set a particularly difficult problem for classification because neither the parameters of the different contagion mechanisms are known, nor any ground truth is available for validation of the classification results. In the following, we propose pathways that yet allow us to learn about the distinguishability of contagion mechanisms in the Twitter dataset.

As a starting point, we applied our classifiers designed for Experiment 3, where we have no information about the adoption parameters. Table \ref{table} shows that the two methods give rather unbalanced results, with the random forest detecting large number of spontaneous adoptions and the likelihood approach being biased towards simple contagion. This discrepancy in the results suggests that one or both of the models might not be capturing the interaction patterns within the Twitter data sufficiently well.

When it comes to empirical adoption data collected via social media, one of the largest bias is induced by the waiting time $t_w$ \cite{karsai2016local}, that is the time gap between the moment someone becomes convinced by an idea (upon exposure) and the moment we can actually observe it through an active adoption event (posting). We report the waiting time distribution for the Twitter dataset in Figure \ref{ad_statistics}a, where one time step corresponds to time span between two consecutive tweets. This $t_w=t_a-t_e$ lag between the exposure $t_e$ and the adoption $t_a$ time can depend on individual user characteristics. It biases our observations as during this $t_w$ time further exposures can appear, that in principle could not be even necessary for the subsequent adoption (``incubation''). Nevertheless, the only observation we can make is about the sequence of influencing tweets, as we can not know the exact tweet that triggered the adoption. The effects of such biases have been studied earlier in other scenarios of online adoption~\cite{karsai2016local,wang2018effects}.
In light of these observations, it is clear from the likelihood computations and from the feature importance ranks shown in Figure \ref{best_features} that both the approaches used so far are ill-suited in this case since they heavily rely on precise adoption times ---assuming no waiting time. To steer our classification algorithms away from making estimates based on this hard assumption, we now introduce a synthetic contagion process evolving on an activity driven temporal network model parameterised from data, and where waiting times can be measured. 


\subsubsection*{Activity driven networks with asynchronous dynamics}

\begin{figure*}[h!]
\centering
\includegraphics[width=160mm]{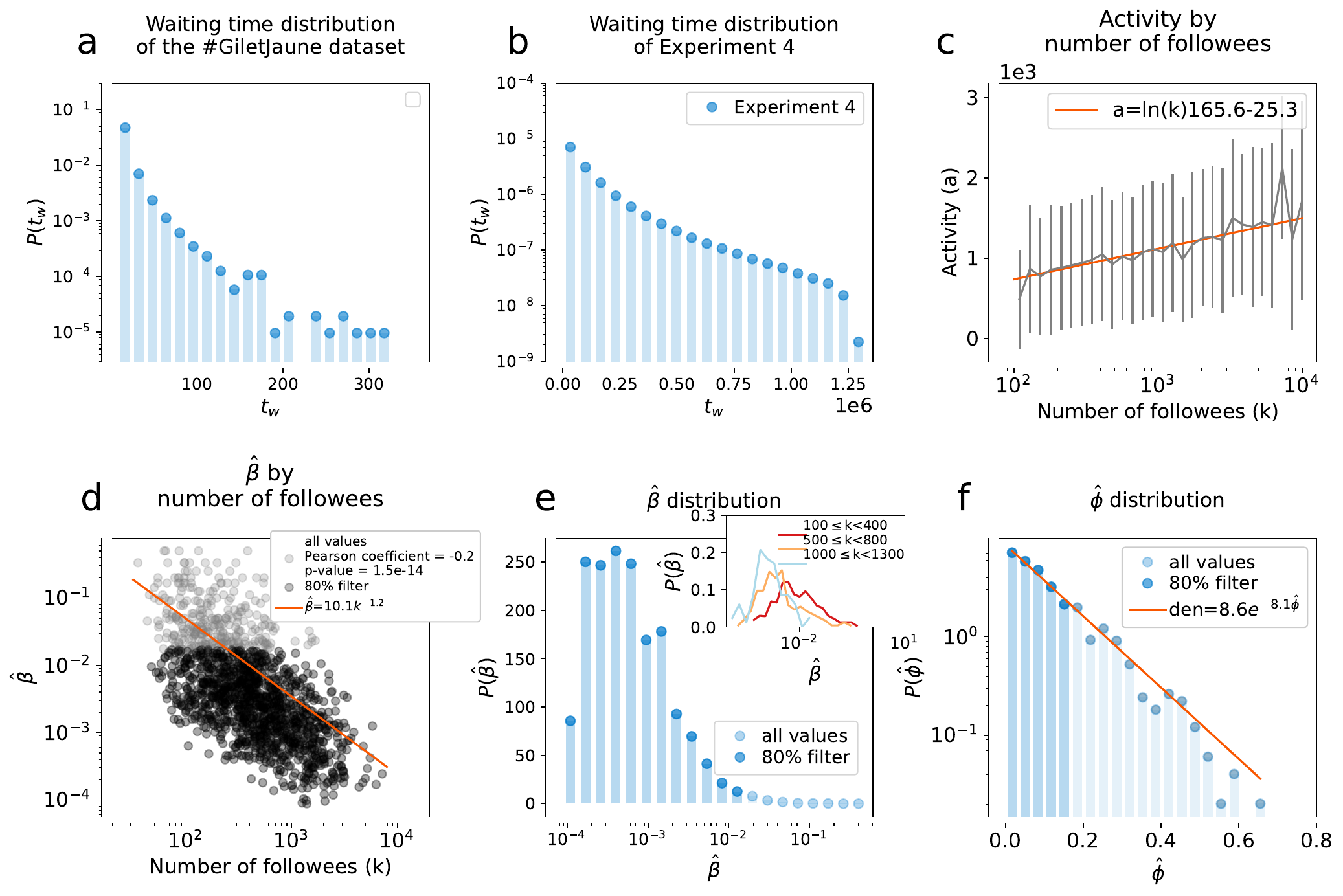}
\caption{Parameter distributions and dependencies of Experiment 4 inferred from the \#GiletsJaunes Twitter dataset. The waiting time distributions observed in the (a) \#GiletsJaunes dataset and (b) in Experiment 4. (c) Correlation between the activities and degrees of nodes in the synthetic propagation inferred from the distribution of the number of tweets posted during the week before adoption as function of the number of active followees in the Twitter dataset. (d) 
Correlation between the inferred simple contagion parameter $\hat{\beta} = 1/(\text{number stimuli})$ and node degrees observed for egos in the \#GiletsJaunes dataset. (e) Distribution of the inferred simple contagion parameter $\hat{\beta}$. The inset depicts the same distribution stratified by degree. (f) Distribution of the $\hat{\phi}$ complex contagion parameter inferred as the proportion of infected neighbours at the time of adoption of an ego in the \#GiletsJaunes dataset ($\hat{\phi}$).
Since the $P(\hat{\beta})$ and $P(\hat{\phi})$ are broad, we apply a filter to retain the $80\%$ of their smallest values.}
\label{ad_statistics}
\end{figure*}


\paragraph{Experiment 4.} We employ a connected and undirected sample of the follower Twitter network as the underlying structure for the contagion process (for more details about the network creation see section Methods, Experiments). We assume that nodes can be in three distinct states: susceptible (not yet infected), aware (they are already infected, but that has not been observed yet through an active post), and detected (they are infected and this has been observed). Every node $i$ is assigned with an activity $\hat{a}_i\in [0,1]$ sampled from a normal distribution with an average activity that characterises nodes coming from the same degree group as node $i$ (Figure \ref{ad_statistics}c). Further, nodes are endowed with parameters $\hat{\beta}_i$ and $\hat{\phi}_i$ respectively sampled from the empirical distributions $P(\hat{\beta})$ and $P(\hat{\phi})$ shown in Figure \ref{ad_statistics}d,f.
Since these distributions are broad, we filtered them and kept only samples from their lowest $80\%$ (more details about sampling and filtering in the Methods section and Supplementary Material 5).

At every step, a node is selected with a probability proportional to its activity, modelling its action of posting. If the selected node is susceptible, we assume its post induces no influence on its neighbours. Once a node is infected via one of the considered mechanisms, it enters the aware state and no further stimuli are necessary for adoption ---yet to be observed. The next time the node is selected for an interaction, it becomes detected. If a node is aware or detected, its posts are considered as influencing events to its neighbours. The resulting waiting time, measured for each infected node as the time between the aware and detected state, follows a broad distribution (Figure \ref{ad_statistics}b), similar to the empirical observations. More details about the model definition and evaluation are explained in Methods. 

The complexity of Experiment 4 makes the application of the Bayesian likelihood method unfeasible, so we continue our investigation only through the random forest approach, using the same feature set as in the previous experiments, and assuming unknown contagion parameters. As before, we pre-assign an adoption mechanism to each node in the modelled activity driven network and compute the classification accuracy. Results, shown in Figure \ref{accuracies_methods}g, demonstrate that despite the increased complexity of this data-driven experiment, the random forest can achieve good classification accuracy all across the parameter space, with average accuracy $0.72$.
In this experiment, the spontaneous adoptions are the hardest to classify since they appear with a very low rate (see the Supplementary Material, Table \ref{accuracy_st_exp4}). 
It is worth noticing that the importance of the features is different from the one previously shown for Experiment 2 (Figure \ref{best_features}). While the feature \emph{time since the last infected neighbours} diminishes in importance due to the presence of a waiting time, the \emph{proportion of the infected neighbours}, and particularly the \emph{degree of the central ego} gain significance (Figure \ref{best_features}).

\subsubsection*{Classification of Twitter hashtags}

\begin{figure*}[h!]
\centering
\includegraphics[width=170mm]{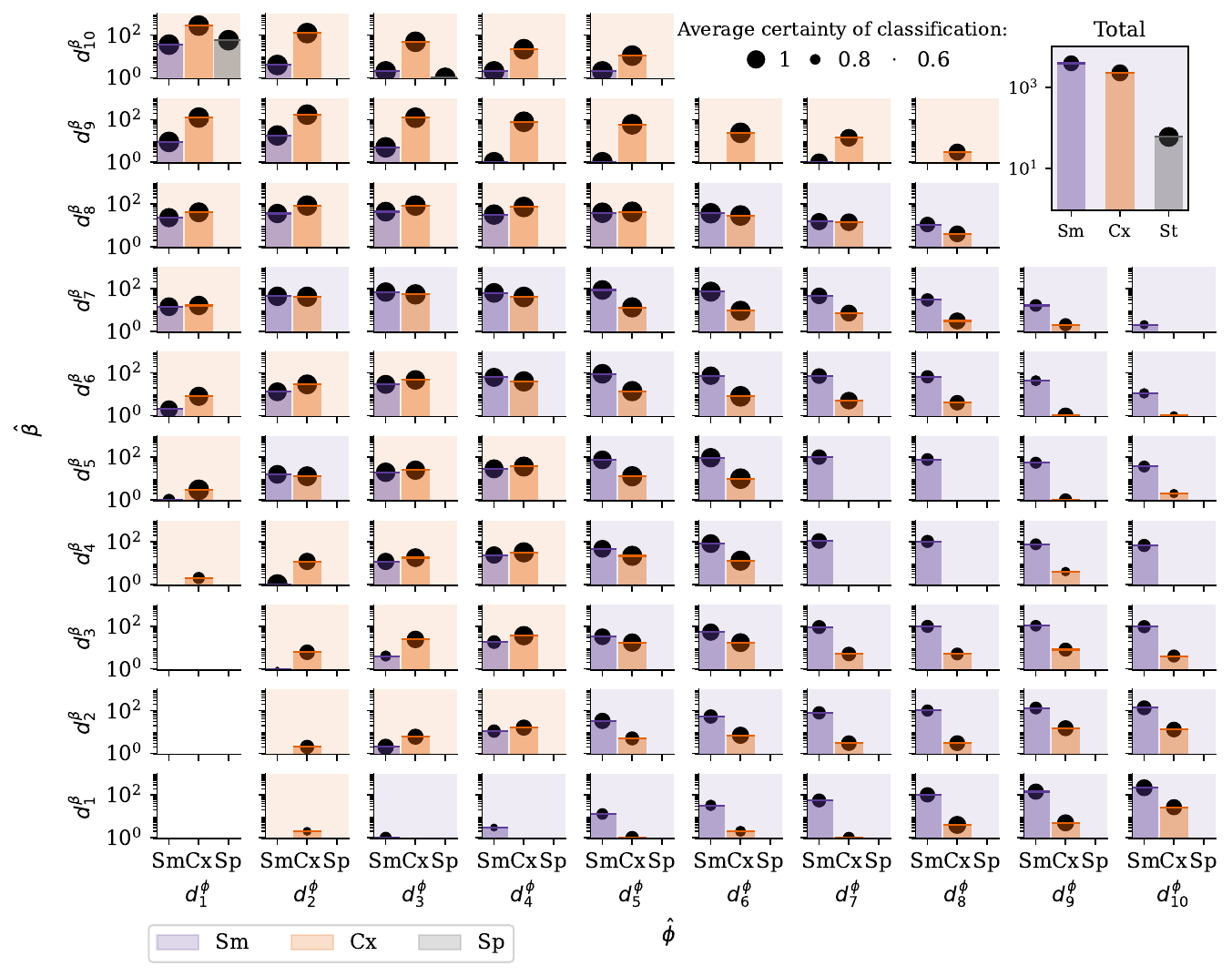}
\caption{Classification of contagion mechanisms of the \#GiletsJaunes Twitter dataset as the function of $\hat{\phi}$ (x-axis) and $\hat{\beta}$ (y-axis) parameters. The notation $d^{parameter}_n$ represents the $n^{th}$ deciles of the parameter distribution from the \#GiletsJaunes dataset from Figure \ref{ad_statistics}. The classification results of each instance $i$ are shown at the corresponding location of the decile of its inferred $\hat{\phi_i}$ and $\hat{\beta_i}$ parameters sampled from the $P(\hat{\phi})$ and $P(\hat{\beta})$ distributions.
The background colour of each panel indicates the dominating classified mechanism that characterise the given parameters (purple for Sm, orange for Cx and blue for Sp). The certainty of classification, displayed with black circles, defined as the proportion of trees in the random forest that have classified an instance into the assigned contagion type, averaged over the set of instances classified in that contagion type. Most of the infection cases are classified as simple if their $\hat{\beta}$ are in the $8^{th}$ decile or below and their proportion of infected neighbours is greater than $d_5^\phi$, and as complex otherwise.}
\label{fig_gj}
\end{figure*}

\paragraph{Experiment 5.} 
To conclude our case study on the Twitter dataset, we apply the trained models from Experiment 4 on the adoption cases of \#GiletsJaunes and related hashtags. The inset of Figure~\ref{fig_gj} shows that most adoption cases are classified as simple as opposed to complex. This suggests that more people adopt \#GiletsJaunes through a repeated influence from their contacts than through combined influence mechanisms. The less detected class is the one of spontaneous adoptions, suggesting the limited influence of external sources with respect to peer-induced contagion within the platform.

Since no ground truth exists for this dataset, instead of visualizing the accuraciy values on the $(\beta,\phi)$ phase space, we show in Figure \ref{fig_gj} the full distribution of inferred adoption mechanisms stratified by their inferred contagion parameters $\hat{\beta}$ and $\hat{\phi}$ (aggregated in deciles). We can see that ego nodes with high $\hat{\beta}$ and low $\hat{\phi}$ values are more likely to be classified as Cx, whereas egos with low $\hat{\beta}$ and high $\hat{\phi}$ tend to be classified as Sm. However, Figure \ref{fig_gj} also suggests that the two inferred parameters, $\hat{\beta}$ and $\hat{\phi}$, cannot capture the complexity of the classification problem on their own. Indeed, both Sm and Cx adoptions appear throughout the parameter space, highlighting the added value of the random forest classifiers trained in our modelling framework. Finally, we observe that certainty of the classification algorithm improves with lower $\hat{\beta}$ and higher $\hat{\phi}$ values, which can be explained by the increased number of stimuli, and therefore a richer dataset, in this parameter range.

%

\section*{Discussion}

Our goal in this work was to infer social contagion mechanisms leading to the adoptions of products, ideas, information, or behaviours. We restricted the focus to three complementary contagion mechanisms potentially determining the behaviour of an ego node, whether adopting spontaneously (exogenous influence) or due to transmission on a social network (endogenous influence) via simple or complex contagion mechanisms.
The general problem of distinguishing social contagion mechanisms in networked populations has recently been addressed by analysing macroscopic spreading curves at the population level~\cite{hebert2020macroscopic,cencetti2023distinguishing,monsted2017evidence}, typically assuming that only one a single mechanisms is exclusively present during the contagion process.
In this work, we overcome these assumptions by ({\it i}) considering only microscopic information at the level of the adopter and their peers and ({\it ii}) allowing different contagion mechanisms to be simultaneously present ---with different parameters--- during the same spreading phenomenon. Under these assumptions, we tackled the inference question as a classification problem under a Bayesian likelihood and a random forest approach over a sequence of experiments with increasing levels of complexity. 
We showed, in controlled synthetic settings, that the limited information available from an ego and its peers is generally enough to distinguish the specific adoption dynamics with varying levels of accuracy depending on the contagion parameters. The lines between the mechanisms becomes more blurred in cases when one infectious neighbour is enough to induce the adoption of an ego. This can happen for strongly infectious items spreading via simple contagion (akin to high individual susceptibility) or low individual thresholds in adoptions triggered by complex contagion, both cases leading to an immediate local transmission and rapid global spreading.
Interestingly, in the simplest experiments performed via simulations on synthetic static networks, we found little impact of the network structure on the accuracy of the classification task. Recent results have shown that simple contagion leads to similar infection patterns across different network models, while the patterns associated to complex contagion mechanisms are less robust~\cite{contreras2024infection}. This could explain the fact in Experiment 2 and 3 we do not observe major differences in the distinguishability of the mechanisms over different network structures, from Erd\H{o}s-R\'enyi graphs to those generated via Barab\'asi-Albert, Watts-Strogatz, and Stochastic Block Model approaches. 
Increasing the level of realism, we demonstrated that simplistic models fail to capture the full complexity proper of real-world transmissions, such as waiting times, or the non-static structure of empirical social networks. The challenges arising in these scenarios confirm the inherent difficulty that comes with these tasks when several internal and external factors are at play at both the dynamical and structural level, as also highlighted in other recent studies that tackled the inference problem in different contexts~\cite{st2024nonlinear}. Nevertheless, even in these realistic settings when mechanistic approaches seem to be out of reach, a random forest classifier trained on a carefully parametrized synthetic model can give interpretable results. 

Despite the comprehensive approach to the inference problem in this paper, our results presented here have certain limitations. First, for simplicity reasons we only consider static network structures, while in reality social influence is mediated via temporal interactions.
Further, we assumed that the effects of external influence (like advertisements and news) do not vary in time, that is clearly an approximation.
During our likelihood formulation we assume each contagion instance to be independent from each other, which is only an approximation, that is accounted for in the random forest approach. Finally, since no real dataset is available with ground truth information regarding the adoption mechanisms of a social contagion, it prevents us to validate our findings in our final experimental setting. Beyond accounting for these limitations, possible extensions of the present method could include the analysis of the spreading of different items on the same population; or to classify different infection mechanisms ~\cite{hodas2014simple, ruan2015kinetics} even beyond pairwise exposures~\cite{iacopini2019simplicial, st2022influential, ferraz2023multistability}, as considered in a recent work~\cite{cencetti2023distinguishing}.

We believe that our results open the door to the investigation of microscopic social contagion mechanisms at the local network level. In one way, our study aims to contribute to the understanding how seemingly similar macroscopic processes can be differentiated at the microscopic level. In another way, we hope to lay down a path to study social contagion processes at the level of individuals, that is more feasible from a real data perspective and can lead us to a more fine-grained understanding how local decision mechanisms lead to system level global phenomena in social contagion processes.

\section*{Methods}

\subsection*{Experiments}

To study the distinguishability of the Sm, Cx and Sp contagion processes we defined three experimental settings with increasing complexity:

\subsubsection*{Experiment 1 - classification on egocentric networks}

In Experiment 1 we assume no underlying network structure to disseminate the spreading process but we operated only with isolated ego networks. We assume knowledge only about egos and their neighbours, that together defined a star structure around the central ego. The degrees of the ego (i.e. number of its neighbours) are drawn from a binomial distribution of parameters $(N, p) = (1000, 0.004)$ (which yields a mean of $\langle k\rangle=4$), excluding the value 0. This was necessary to obtain the same parametrization than the Erdős–Rényi networks that we used in Experiment 2. We assign to each ego-node a predetermined adoption class, simple or complex, with corresponding parameter, respectively $\beta$ or $\phi$. Further, we defined the same adoption probability $r_{nb}$ for any neighbour of an ego, mimicking their adoption dynamics as a Bernoulli process. Assuming each node in the ego-network to be susceptible at the outset, neighbours became infected following their Bernoulli dynamics, while egos changed state only when their condition to infect has been satisfied. We simulate this contagion dynamics on $100,000$ ego-networks, having $10,000$ realisations for each parameter values of $\beta$ and $\phi$ taking values from \{0.1, 0.3, 0.5, 0.7, 0.9\} and with parameter $r_{nb}=0.05$. In this setting the classifier was informed by the $\beta_i$, $\phi_i$ and $r$ parameter values for each instance $i$.

\subsubsection*{Experiment 2 - classification on random networks with known parameters}

Experiment 2 is conducted on an Erdős–Rényi model network \cite{erdHos1960evolution}, with $1,000$ nodes and average degree $4$. For comparison purposes, in Supplementary Material 4, we also demonstrate our results using Watts-Strogatz \cite{watts1998collective} and Barabási–Albert \cite{barabasi1999emergence} model networks, Stochastic Block model networks \cite{lee2019review}, and a real Twitter mention network \cite{unicomb2019reentrant} defined by linked customers if they mutually mentioned each other during the observation period. For computational purpose we filter the Twitter mention network to keep only its largest connected component, i.e. the largest interconnected subset of nodes within a network (370,544 nodes and 1,013,096 links) and we assume it to be undirected by ignoring the directions of its links. 
As in Experiment 1, we assign all nodes beforehand with a contagion process (Sm or Cx) and a parameter ($\beta$ or $\phi$) accordingly from the set \{0.1, 0.3, 0.5, 0.7, 0.9\} in order to have all pairs $(process, parameter)$ equally distributed in the data set. Having all nodes as susceptible at the outset, the propagation initialised by infecting one random node. The spreading process among the rest of the nodes is gradually spreading either by their assigned process of contagion, or through the spontaneous adoption with a rate of $r$.
We stop the contagion process when all of the nodes become infected, except for the Twitter mention network, where the process is terminated when 90\% of the nodes become infected. 
For each synthetic network model, the propagation is run on 20 independent network realisations, with $r=0.005$. For each node $i$, the parameters $\beta_i$, $\phi_i$ and $r$ are assumed to be know by the classifiers.

\subsubsection*{Experiment 3 - classification on random networks with unknown parameters}
Experiment 3 is aiming to solve the classification of the same contagion instances than Experiment 2 but without prior knowledge about the parameters of $\beta_i$, $\phi_i$ and $r$.

\subsubsection*{Experiment 4 - classification on real networks with known parameters}

Experiment 4 is inspired by the Activity Driven network model \cite{perra2012activity} and has been created to represent the propagation of a hashtag on the Twitter platform. Here we use the largest connected component of an un-directed mutual follower network from Twitter \cite{unicomb2019reentrant} and concentrate on the propagation of the hashtags related to the political movement called \#GiletsJaunes. For computational purposes, we iteratively filter this network to reduce its size. At the outset, the filtered network only contains one randomly selected node from the initial network. Subsequently, a neighbour of the initial node is selected with a probability inversely proportional the node's degree. Once a neighbour is selected, it is incorporated into the filtered network along with its edge. Subsequently we reproduce this process, each time selecting a neighbour from the newly integrated node and its edge, until we achieve a network size of 100,000 nodes.

\paragraph{Parameter sampling.} First of all, in this setting each node is assigned with an activity, mimicking its level of participation on the Twitter platform. As the distribution of the number of tweets posted by each user during a week depends on its degree and because those distributions along a certain degree range are not part of the typical known distributions, we sample the assigned activity of each node with a normal distribution centred on the average number of tweets posted by each user corresponding to its degree.

Further, we sample homogeneously each node to assign them with an adoption process, being simple or complex contagion.

Further, parameters are sampled for each node depending on the assigned mechanism. For simple contagion parameter values for $\hat{\beta}$ are defined as the inverse of the number of times a hashtag appeared in the timeline of an observed ego's neighbours, one week before the ego's adoption. Note that we consider cases of infected egos who have at least one infected neighbour at the time of adoption. Since the $\hat{\beta}$ parameter shows correlation with the node degree (see Figure \ref{ad_statistics}d), we decided to account for this dependency when sampling $\hat{\beta}$ values for egos. We group nodes by their degrees and assume that each $P(\hat{\beta})_k$ distribution for a degree class can be approximated by a log-normal distribution with an average characterising the actual degree class (see Figure~\ref{ad_statistics}e and its inset). Thus for each node $i$ with degree $k$ to obtain a $\hat{\beta}_i$ we simply sample the corresponding log-normal distribution.

At the same time, the parameter $\hat{\phi}_i$ for the complex contagion mechanism is measured as the fraction of infected neighbours of an ego that adopted a hashtag. The distribution of $P(\hat{\phi})$ (in Figure \ref{ad_statistics}f) is measured from adoption cases where the last infected neighbour of the ego before its adoption was a newly infected neighbour. We assign a parameter $\hat{\phi}_i$ to a node $i$ by sampling this distribution $P(\hat{\phi})$ shown in Figure \ref{ad_statistics}f.
Finally, to avoid the sampling of extreme values, since the distributions $P(\hat{\beta})$ and $P(\hat{\phi})$ appeared as broad distributions, we filter them by keeping 80\% of their lowest values for parameter sampling. For a robustness analysis on the effect of filtered fraction of inferred parameters see the Supplementary Material.

\paragraph{Contagion model with waiting time.} Beyond the realistic data-driven parametrisation of the network and adoption mechanisms, our main goal with this experiment is to simulate spreading scenarios to study the effects of waiting times between node adoption and its observation on the inference of spreading mechanisms. For this reason we assume that every node of the network can be in one of the three following states: susceptible (not infected), aware (infected but the infection cannot be observed) and detected (the infection can be observed). After infecting a uniformly randomly selected seed node to launch the spreading process, we iteratively execute the following protocol at each time step: first a node is selected randomly with a probability proportional to its activity, indicating that this node posts a tweet. If the node is susceptible, it can become adopted with probability $r$, mimicking the possibility to post the hashtags spontaneously. Otherwise the susceptible node can get infected through its assigned adoption mechanism. If a node is active but susceptible, its post will not count towards the influence of its neighbour. However, if the node is aware, at the time of its next post it becomes detected. Once aware or detected we assume that at each future activity of a node it will post the spreading hashtag. If a post of a node includes the hashtag, it counts as a stimuli to all of its neighbours, which can become aware if they are susceptible and their condition of infection is reached. In our simulations we modelled the contagion processes in the network until they reached the 90\% of the nodes and used the observed adoption instances for the training of a random forest classifier that was not aware of the contagion parameters.

%


\subsubsection*{Experiment 5 - Classification of hashtag adoption with unknown ground truth}

Since we cannot obtain the contagion mechanisms as ground truth labels for real data, we use the trained model in Experiment 4 for classification of the empirical adoption instances. With these model, that has been trained on data-driven model data closest to reality, we explore the whole $(\hat{\beta},\hat{\phi})$ parameter space, as shown in Figure~\ref{fig_gj}.
We then use this algorithm to analyse various social contagion processes by focusing on tweets with specific sets of hashtags corresponding to distinct topics. We use a dataset collected by \cite{de2020information} from Twitter, now known as X, a social media platform where users can follow each other and share brief posts, or tweets. This dataset spans from May 1, 2018, to May 31, 2019, and includes all tweets from 8,527 selected users interested in the European election of 2019 (denoted as egos) and the accounts they follow (denoted as followees). In total, this comprises 1,844,978 timelines, representing the chronological history of tweets from these users. These tweets cover a range of topics, referenced with key words called hashtags.

Among all the possible hashtags, we choose to focus on \#GiletsJaunes and its variations (\#GiletsJaunes, \#giletsjaunes, \#Giletsjaunes, \#GiletJaune, \#Giletjaune, \#giletjaune, \#giletsjaune, \#Giletsjaune, \#GJ), which are linked to a political movement in France that causes social contagion on Twitter. Our first step is to identify users who have adopted one of these hashtags and then examine the tweets from their followees. We consider that individuals remember influences only from their recent past, thus we study tweets of the ego and its followees on the last week before the adoption. We define a user's degree by the number of followees who posted at least one tweet in the preceding week and we use event time instead of real time for our analysis. Event time counts the number of tweets by followees, regardless of whether they contain the hashtag. We define a stimulus as a tweet posted by followee containing the hashtag.

\subsection*{Likelihood calculations}

The classification with the Bayesian approach follows the same protocol for each experiment: we first compute the likelihood that a given observed adoption case has been caused by each mechanism, being simple, complex or spontaneous, and then we classify the adoption event into the category which maximises the likelihood.

\subsubsection*{Experiment 1}

We determine the likelihood that a node $i$ has been infected either through simple or complex contagion using Eq.~\eqref{llh_tiago}, which expresses the likelihood of the entire process as a product of the likelihoods of each time step (Markov property). We call $\sigma_{i}(t)$ the state of a node $i$ at time $t$, being 0 (S) or 1 (I). To compute the likelihood of observing the ego's state $\sigma_{i}(t+1)$ conditioned on its state and the states of the neighbours $\sigma_{i, nb}(t)$ in the previous timestep, we distinguish three cases:
\begin{enumerate}
    \item ego stays susceptible, formally $\sigma_{i}(t+1)=\sigma_{i}(t)=0$, which we abbreviate as $0 \rightarrow 0$
    \item ego becomes infected, formally $\sigma_{i}(t+1)=1$, $\sigma_{i}(t)=0$, which we abbreviate as $0 \rightarrow 1$
    \item ego stays infected, formally $\sigma_{i}(t+1)=\sigma_{i}(t)=1$, which we abbreviate as $1 \rightarrow 1$.
\end{enumerate}
In case of a simple contagion, the independence of infection probabilities on each edge makes it possible to combine the three cases into a single equation as
\begin{equation*}
\displaystyle
  \mathcal{L}(\sigma_{i}(t+1)|\sigma_{i, nb}(t), Sm, \beta) =
    \begin{cases}
      \displaystyle \prod_{j \in nb}(1-\beta)^{\sigma_j (t)} & 0\rightarrow0\\
      1-\displaystyle \prod_{j \in nb}(1-\beta)^{\sigma_j(t)}   & 0\rightarrow1\\
      1 & 1\rightarrow1
    \end{cases}       
\end{equation*}
where $nb$ is the set of the neighbours of the ego.

In case of a complex contagion, the same likelihood function takes the binary values
\begin{equation*}
\displaystyle
  \mathcal{L}(\sigma_{i}(t+1)|\sigma_{i, nb}(t),Cx, \phi) =
    \begin{cases}
      \mathbb{1}(\sigma_{i, nb}(t)) & 0\rightarrow0\\
      1-\mathbb{1}(\sigma_{i, nb}(t))   & 0\rightarrow1\\
      1 & 1\rightarrow1
    \end{cases}       
\end{equation*}
depending on whether the condition
\begin{equation*}
\mathbb{1}(\sigma_{i, nb}(t))=\Theta\left(\displaystyle\sum_{j} \sigma_j(t) A_{ij} - \phi \displaystyle\sum_{j} A_{ij}\right),    
\end{equation*}
on the proportion of infected nodes is satisfied or not. In this case $A$ denotes the adjacency matrix of the network, with elements $A_{ij}$, and $\Theta$ denotes the Heaviside step function, which is equal to 1 if the input if positive, 0 otherwise.

\subsubsection*{Accuracy estimation for Experiment 1}

In Experiment 1, the accuracies from of the maximum likelihood classification algorithm can be computed analytically across the phase space. Let us define $\hat{X}$ to be the contagion label that the algorithm assigns, and $X$ to be the true contagion label. Assuming a uniform prior on the contagion labels, the accuracy of the algorithm can be expressed as: 
$$\frac{P(\hat{X} = Cx \mid X = Cx) +  P(\hat{X} = Sm \mid X = Sm) }{2}.$$
Since for a node infected by the complex contagion, we always have 
$\mathcal{L}(\sigma_{i}(t+1)|\sigma_{i, nb}(t),Cx, \phi)=1$, the maximum likelihood approach always classifies complex nodes correctly. Consequently, $P(\hat{X} = Cx \mid X = Cx) = 1$ always holds. 

For the second term, to compute 
$$P(\hat{X} = Sm \mid X = Sm)=1 - P(\hat{X} = Cx \mid X = Sm),$$
we need to estimate the probability that a node $i$ with degree $k$ becomes infected by the simple contagion immediately after $\lceil k\phi \rceil$ of its neighbours get spontaneously infected, and therefore it incorrectly becomes classified as complex. Conditioning on the event that the ego has $n$ infected neighbours at time $t$, we define following two random variables:
\begin{itemize}
    \item $N_n$ denotes the number of time steps until a new neighbour gets infected
    \item $E_n$ denotes the number of time steps until the ego gets infected, assuming that no new neighbor gets infected.
\end{itemize}
Since at each time step, the probability of a new neighbour spontaneously becoming infected is $p_n=1-(1-r)^{k-n}$, the random variable $N_n$ follows a geometric distribution with success probability $p_n$. Similarly, since the probability that any of the $n$ neighbours infect the ego node in each time step is $b_n=1-(1-\beta)^{n}$, the random variable $E_n$ follows a geometric distribution with success probability $b_n$. Our goal is to compute the probability of the event that the ego becomes infected immediately after $\lceil k\phi \rceil$ of its neighbours get infected, i.e. that $N_n<E_n$ holds for $n<\lceil k\phi \rceil$, but $E_{\lceil k\phi \rceil}=1$. For each $n<\lceil k\phi \rceil$, the corresponding event probability can be computed based on the well-known formula of two competing geometric random variables. For $n=\lceil k\phi \rceil$, the event probability is simply $b_n$. Finally, due to the Markov property of the contagion process, assuming that no two neighbours get infected at the same time, we arrive to the final result by computing the product of the event probabilities for each $n$:
\begin{equation*}
    \begin{aligned}
    P(\hat{X} = Cx \mid X = Sm) 
    & \approx \left(\prod_{n=1}^{\lfloor k\phi \rfloor} \frac{p_n-p_nb_n}{b_n+p_n-p_nb_n} \right) b_{\lceil k\phi \rceil}.
    \end{aligned}
\end{equation*}

Our result is an approximation, because we did not account for the low-probability event that two neighbours might be infected at the same time. Despite this limitation, the outcomes closely align with the accuracy values observed in the simulations (see Figure \ref{accuracies_methods}, panel d).

\subsubsection*{Experiment 2 - classification with known parameters}

The calculations of the likelihoods of Experiment 2 are similar to Experiment 1, but instead of two, now they involve three processes: simple, complex and spontaneous adoptions. For clarity, we divide those three processes in four scenarios: \\

1. The ego, initially assigned with the simple contagion, eventually becomes infected by the simple contagion: \\

\begin{align}
\label{scsc}
  \mathcal{L}(\sigma_{i}(t+1) | &\sigma_{i, nb}(t),\text{Sm}, \beta) =\\
    &\begin{cases}
        (1 - r)\displaystyle \prod_{j \in nb}(1-\beta)^{\sigma_j (t)}  & \hspace{-0.2cm} 0\rightarrow0\\
       (1-r) \left( 1-\displaystyle \prod_{j \in nb}(1-\beta)^{\sigma_j(t)} \right)  & \hspace{-0.2cm} 0\rightarrow1\\
      1 &  \hspace{-0.2cm} 1\rightarrow1
    \end{cases}  
\end{align}

2. The ego, initially assigned with the simple contagion, eventually becomes infected by the spontaneous contagion: \\

\begin{align}
  \mathcal{L}(\sigma_{i}(t+1) |&\sigma_{i, nb}(t),\text{Sm }\rightarrow\text{ St}, \beta) = \\
    &\begin{cases}
      (1 - r) \displaystyle \prod_{j \in nb}(1-\beta)^{\sigma_j (t)} & 0\rightarrow0\\
      r \displaystyle \prod_{j \in nb}(1-\beta)^{\sigma_j(t)}   & 0\rightarrow1\\
      1 & 1\rightarrow1
    \end{cases}
    \label{scsp}
\end{align}

3. The ego, initially assigned with the complex contagion, eventually becomes infected by the complex contagion: \\

\begin{align}
  \mathcal{L}(\sigma_{i}(t+1) |&\sigma_{i, nb}(t),\text{Cx}, \beta) =\\
    &\begin{cases}
        (1-r) \left( 1 - \mathbb{1}(\sigma_{i, nb}(t))  \right)  & 0\rightarrow0 \\
        \mathbb{1}(\sigma_{i, nb}(t)) & 0\rightarrow1\\
      1 & 1\rightarrow1
    \end{cases}  
    \label{cccc}
\end{align}

4. The ego, initially assigned with the complex contagion, eventually becomes infected by the spontaneous contagion: \\

\begin{align}
  \mathcal{L}(\sigma_{i}(t+1) | &\sigma_{i, nb}(t),\text{Cx }\rightarrow\text{ St}, \beta) = \\
    &\begin{cases}
        (1-r) \left( 1 - \mathbb{1}(\sigma_{i, nb}(t))  \right)  & 0\rightarrow0\\
        r \left( 1 - \mathbb{1}(\sigma_{i, nb}(t))  \right)& 0\rightarrow1\\
      1 & 1\rightarrow1
    \end{cases}.
    \label{ccsp}
\end{align}

\subsubsection*{Experiment 3 - classification with unknown parameters}

In this case we assume that parameter values are not known for the classifier and we employ the same formulas as in Equations~\eqref{scsc}-\eqref{ccsp} used for classifying contagion instances from Experiment 2 with known parameters. However, here the parameters $\beta$ and $\phi$ and $r$ are no longer the true values but are instead inferred from the modelled spreading process: $\hat{\beta}$ as the inverse of the number of stimuli, $\hat{\phi}$ as the proportion of infected neighbours and $\hat{r}$ as the fraction of time spent by a node in the S state with at least one infected neighbour, averaged on every node in that case.

%
%

\subsection*{Random forest classification}

\subsubsection*{Experiment 2 - classification with known parameters}

We train 25 random forest algorithms, one for each pair of $(\beta, \phi)$ by sampling 18,000 instances from Experiment 2, with 6,000 contagion cases from each category. Then we test the models on a set containing 6,000 instances (2,000 instances from each category). The results are averaged over 10 realisations. Each random forest algorithm has 100 trees without any limit on the maximum of depth. The use of the Gini function or the entropy function is determined by grid search.

\subsubsection*{Experiment 3 - classification with unknown parameters}

We train a unique random forest model on a sample of Experiment 2, which contains 18,000 instances in total (6,000 instances in each category), regardless of the parameters. The results are averaged over 10 realisations. Each random forest algorithm has 100 trees without any limit on the maximum of depth. The use of the Gini function or the entropy function is determined by grid search.

%
%
%
%
%
%



\section*{Acknowledgements}

The authors are thankful for T. Peixoto for the constructive discussions and for S. Centellegher for releasing visualisation tools used in this paper.
G.Ó. was supported by the Swiss National Science Foundation, under grant number P500PT-211129. I.I. acknowledges support from the James S. McDonnell Foundation's $21^{\text{st}}$ Century Science Initiative. M.K. acknowledges funding from the National Laboratory for Health Security (RRF-2.3.1-21-2022-00006), the ANR project DATAREDUX (ANR-19-CE46-0008); the SoBigData++ H2020-871042; and the MOMA WWTF project.

\section*{Author contributions statement}

All authors contributed to the development of the research design. E.A. performed the numerical simulations and data analysis. E.A., G.Ó. and I.I. developed the statistical analysis. All authors wrote the first draft of the manuscript, interpreted the results, and edited and approved the manuscript.

\section*{Competing interests} The authors declare that they have no competing interests. The funders had no role in study design, data collection and analysis, decision to publish, or preparation of the manuscript.

%
%
%
\bibliography{main}

\clearpage
\newpage

\onecolumn

\setcounter{figure}{0}
\setcounter{table}{0}
\setcounter{equation}{0}
\setcounter{section}{0}

\makeatletter
\renewcommand{\thefigure}{S\arabic{figure}}
\renewcommand{\theequation}{S\arabic{equation}}
\renewcommand{\thetable}{S\arabic{table}}

\setcounter{secnumdepth}{2} 

	\textbf{\Huge Supplementary Material}
 \\
 \\
 
 \textbf{\Large Distinguishing mechanisms of social contagion from local network view}

\section{Contagion curves of the extreme values of the phase space}

\begin{figure}[h!]
\centering
\includegraphics[width=150mm, trim={6cm 6cm 6cm 1.2cm}]{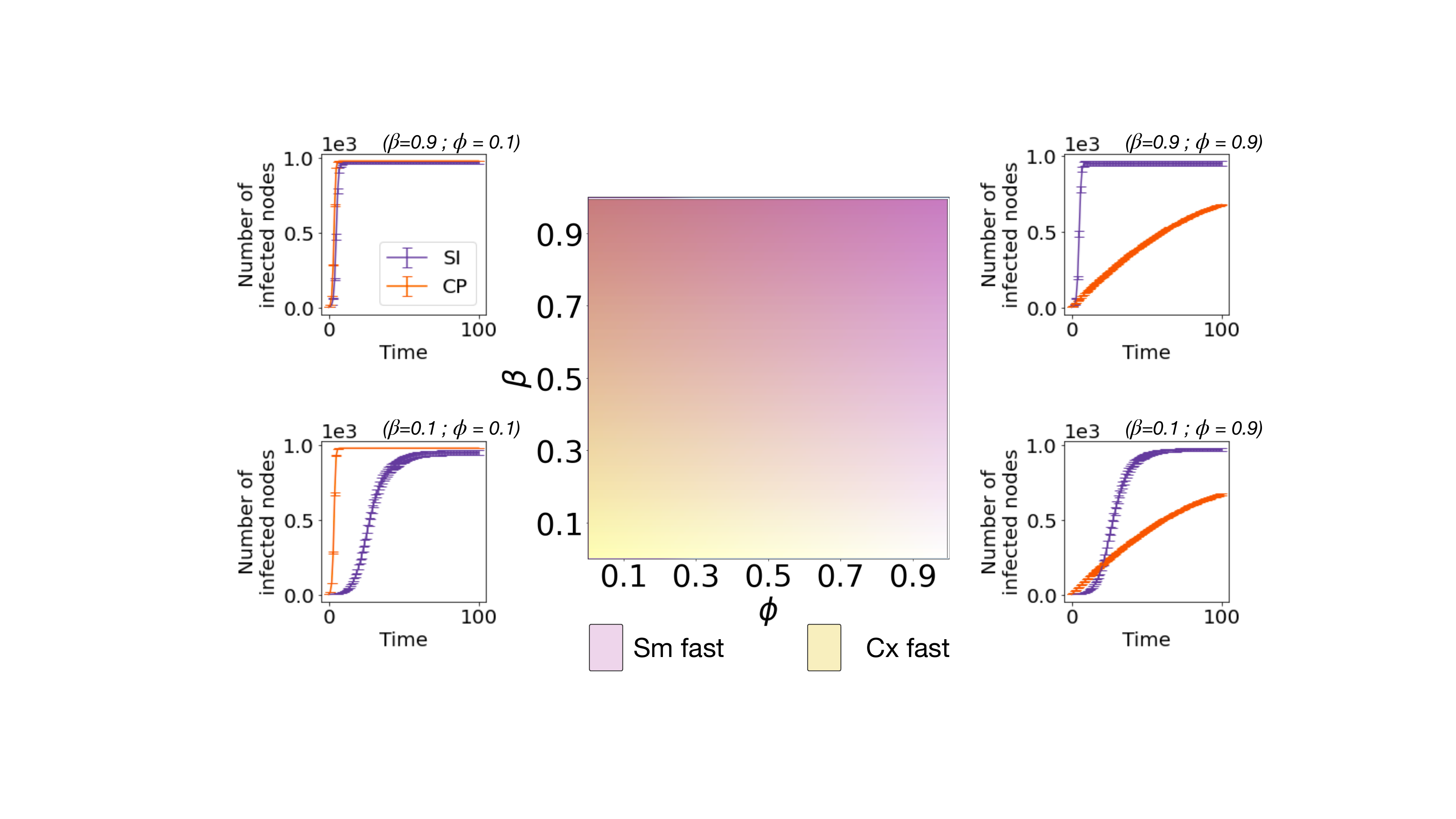}
\caption{(central panel) Speed dependency of the simple and complex contagions on the epidemic parameters $\beta$ and $\phi$ (respectively). The purple and orange colours display respectively the areas where the simple and the complex contagions are faster. The number of infected nodes through time is shown on the four extremes of the parameter space, in purple for the simple contagion and in orange for the complex contagion.}
\label{fig_contagion_curves}
\end{figure}

We investigate the speed of spreading in processes governed by the simple and the complex contagion mechanisms independently. These two dynamical processes are implemented separately on an Erdős–Rényi network of 1000 nodes, with an average degree of 4. We examine the speed of spreading as the function of the simple and complex spreading parameters of $\beta$ or $\phi$ (respectively), taking values from a broad range between 0 and 1 (cf Figure \ref{fig_contagion_curves}).
High values of $\beta$ characterise simple contagion processes with high speed since nodes in this scenario have a higher probability to be infected, commonly after a single stimulus.
The opposite effect characterise complex contagion: if $\phi$ is high, the propagation is slow-downed as the proportion of infected neighbours needed adoption is large.

\section{Selection of the algorithm of machine learning for the classification in Experiment 2}

We investigate to find the best classification machine learning algorithm to distinguish between the simple, complex and spontaneous contagions in Experiment 2. We test the following 9 algorithms: \\

\begin{itemize}
    \item \textit{Naives Bayes \cite{zhang2004optimality}:} algorithm which classifies instances using the Bayes' theorem under the hypothesis that every pair of features are independent.
    \item \textit{K-nearest neighbors (Knn)\cite{mucherino2009k}:} the training instances are displayed in a space of the dimension of the number of features. When classifying a contagion case, it is assigned to the same category as the majority of its closest neighbors in this feature space.
    \item \textit{Perceptron\cite{gallant1990perceptron}:} classifier that learns by iteratively adjusting weights. It utilizes a threshold function to determine the class of the instances based on the dot product of input features and learned weights.
    \item \textit{Support Vector Classification (SVM)\cite{hsu2003practical}:} algorithm that identifies an optimal hyperplane to separate data into different classes by maximizing the margin between the classes.
    \item \textit{Linear Support Vector Classification (Linear SVM)\cite{hsu2003practical}:} variant of Support Vector Classification that specifically employs a linear decision boundary to classify data points into distinct categories.
    \item \textit{Decision tree\cite{song2015decision}:} algorithm that recursively partitions data based on feature attributes to construct a hierarchical tree structure for classification
    \item \textit{Random forest\cite{breiman2001random}:} method which builds multiple decision trees during training and combines their predictions through averaging their results.
    \item \textit{Ada boost\cite{wang2012adaboost}:} boosting algorithm that sequentially trains weak learners by emphasizing misclassified instances in subsequent iterations, and thus build a strong classifier by combining the predictions of these weak learners.
    \item \textit{Gradient boosting\cite{natekin2013gradient}:} boosting algorithm that sequentially trains weak learners giving more weight to the misclassified instances based on gradients of a loss function.
\end{itemize}

We present the mean accuracies over the whole parameter-space in the classification of the instances from Experiment 2 for each machine learning algorithm in Table \ref{table_algo}. 
Among the algorithms displaying the highest accuracies (above 0.82), we opt for the \textit{Random forest} method first due to its significantly faster computation times compared to \textit{SVM}.
Additionally, the \textit{Random forest} algorithm, consisting of an ensemble of decision trees whose outcomes are combined, generally outperforms individual \textit{Decision Tree} methods. Finally, we exclude the Gradient Boosting algorithm due to its limited explainability.

\begin{table}[h!]
\begin{tabular}{|l|l|l|l|l|l|l|l|l|}
\hline
Naives Bayes & Knn  & Perceptron & Linear SVM & SVM  & Decision tree & Random forest & Ada boost & Gradient boosting \\ \hline
0.66         & 0.81 & 0.68       & 0.81       & 0.82 & 0.82          & 0.82          & 0.75      & 0.83              \\ \hline
\end{tabular}
\caption{Average over the whole parameter-space of the accuracies on the classification of the contagion cases from Experiment 2}
\label{table_algo}
\end{table}


\section{Distribution of the features of the random forest of Experiment 2}

\begin{figure}[h!]
\centering
\includegraphics[width=120mm, trim={0cm 1cm 0cm 0cm}]{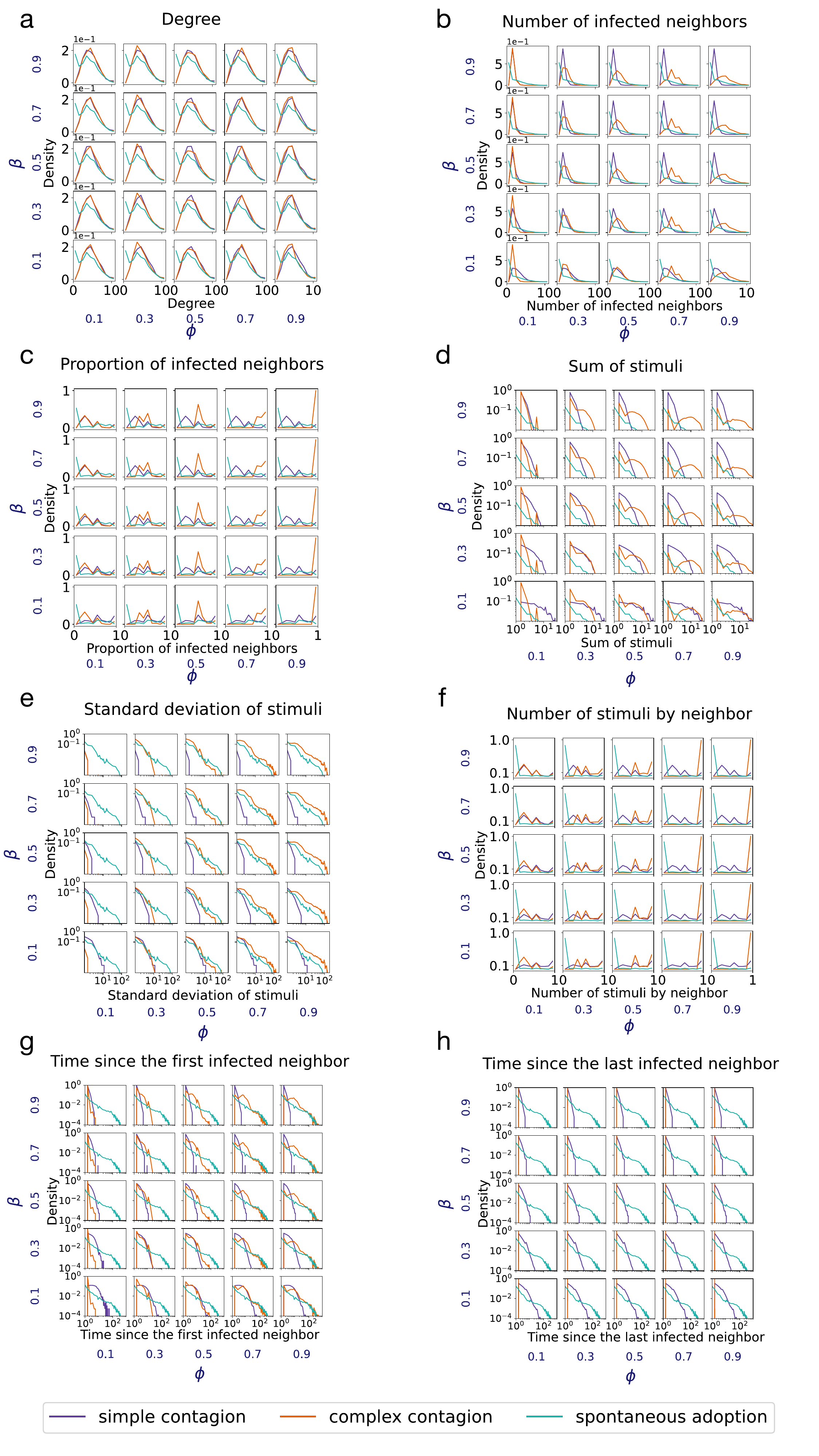}
\caption{Distribution of the features of the random forest algorithms across the parameter space : degree (panel a), number of infected neighbours (panel b), proportion of infected neighbours (panel c), sum of stimuli (panel d), standard deviation of stimuli (panel e), number of stimuli by neighbours (panel f), time since the first infected neighbour (panel g) and time since the last infected neighbour (panel h). The displayed values, taken from Experiment 2, are grouped by their dynamical processes (simple, complex or spontaneous).}
\label{fig_distributions}
\end{figure}

The features of the random forest have been chosen to present different values according to the mechanisms of adoption. As depicted in Figure \ref{fig_distributions}, the distributions of most features differ for the simple, complex and spontaneous adoptions. The degree is the only feature which is not related to the propagation itself but to the structure of the network. While it does not present significant differences in the distributions within Erdős–Rényi networks, we keep it due to the potential influence of a node's degree in other type of networks.

\section{Best subset of features for the random forest on Experiment 2}

\begin{figure}[h!]
\centering
\includegraphics[width=180mm]{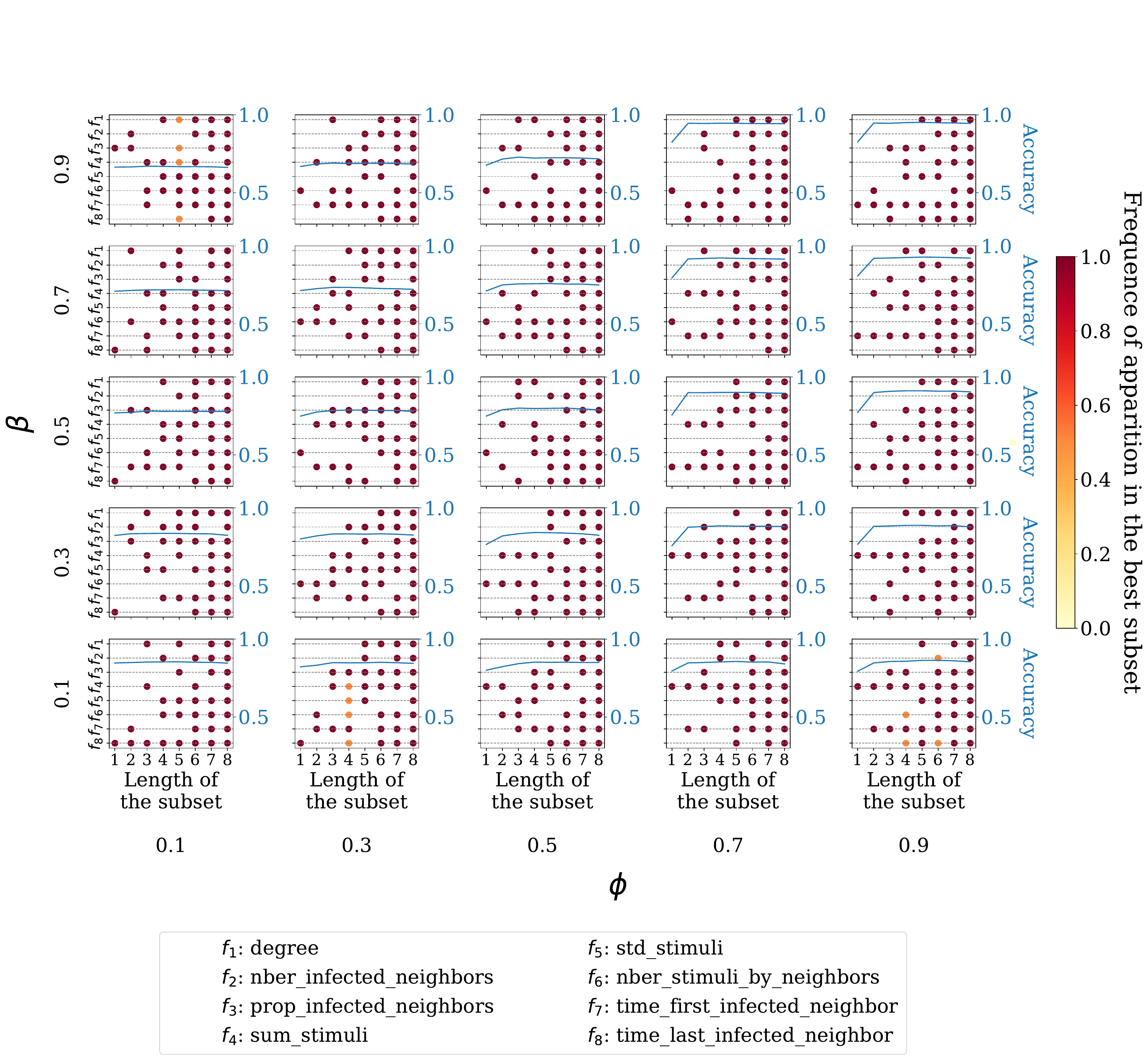}
\caption{Subsets of features giving the best accuracies (y-axis of each subplot) in the parameter space $(\beta, \phi)$ for a certain length of subset (x-axis of each subplot) in the classification with the random forest of Experiment 2. The corresponding accuracies are displayed in blue. If several subsets give the same best accuracies, we compute the frequency of apparition of each feature in those subsets. In most of the cases, only three features is enough to obtain the same accuracy values than with the total set of features, but those three features are different across the parameter space.}
\label{fig_subset_features}
\end{figure}

\begin{figure}[h!]
\centering
\includegraphics[width=180mm]{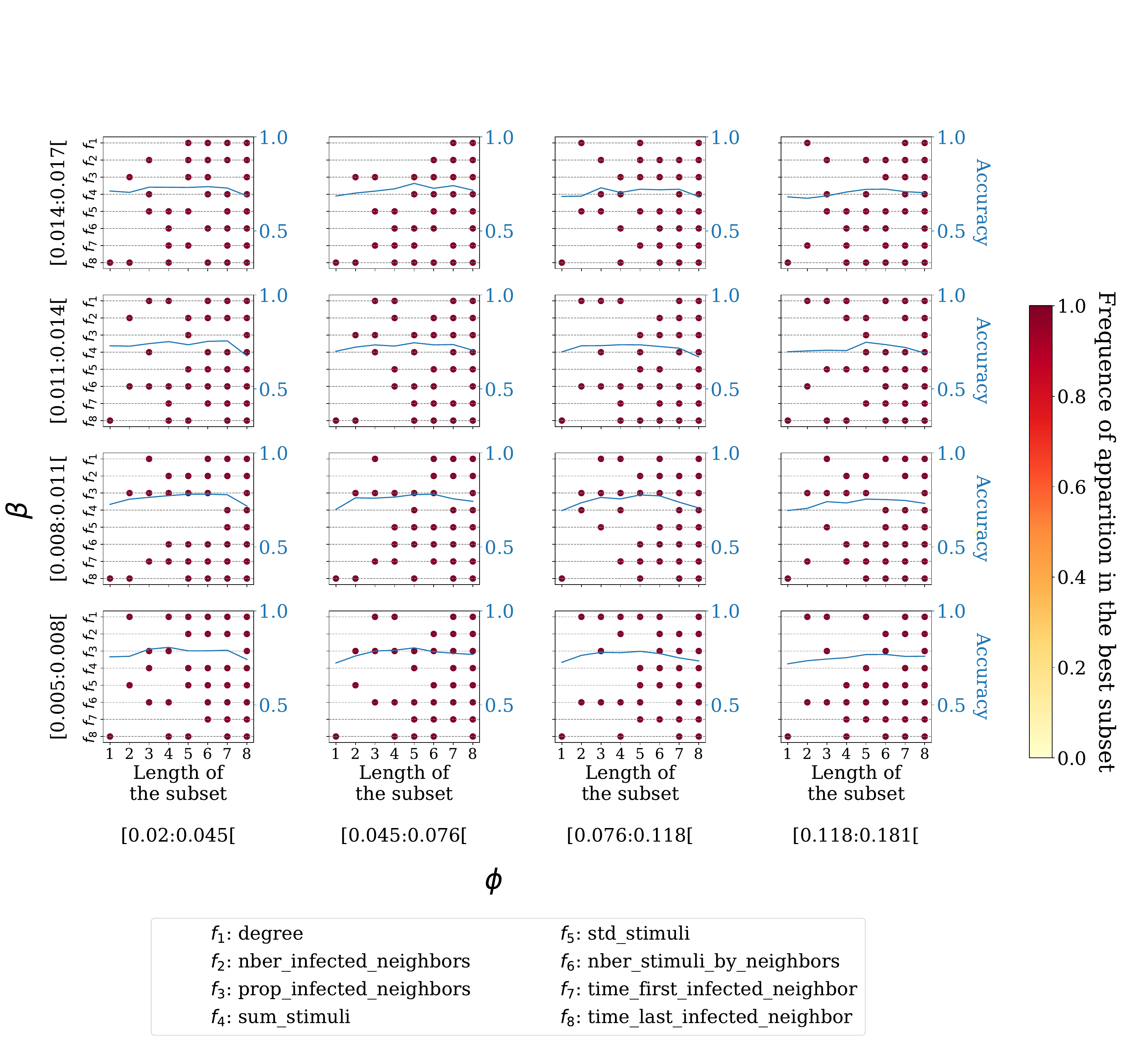}
\caption{Subsets of features giving the best accuracies (y-axis of each subplot) in the parameter space $(\beta, \phi)$ for a certain length of subset (x-axis of each subplot) in the classification with the random forest of Experiment 3. The corresponding accuracies are displayed in blue. If several subsets give the same best accuracies, we compute the frequency of apparition of each feature in those subsets. In most of the cases, only three features is enough to obtain the same accuracy values than with the total set of features, but those three features are different across the parameter space.}
\label{fig_subset_features_ad}
\end{figure}

To evaluate the significance of the features of the random forest method on Experiment 2 and 3, we train algorithms with all possible subsets of the eight features.
Figures \ref{fig_subset_features} and \ref{fig_subset_features_ad} present the subset with the highest accuracy (y-axis) for each subset length (x-axis), respectively for Experiments 2 and 3, across the all parameter-space, with corresponding accuracy values indicated in blue.
Looking at the results from Experiment 2, enlarging the feature set from one to three increases the accuracy, but a plateau is reached for subsets larger than four. 
In other words, in most regions of the parameter space, only three features are necessary to achieve the same accuracy as with more features.
However, this optimal subset varies through the parameter space.
Also, adding features increases the accuracy when $\phi$ is high, but does not have any effect when the value of $\phi$ is small.
Differently, the set of feature for the classification of Experiment 3 does not have a great influence on the accuracies, as the obtained values for different length of subsets are very similar.
In contrast, the selection of features for the classification in Experiment 3 has a limited influence on the accuracies, as the obtained values for different subset lengths are very similar.

\clearpage
\newpage

\section{Accuracies of the different methods on different networks on Experiment 2}

\begin{figure}[h!]
\centering
\includegraphics[width=180mm]{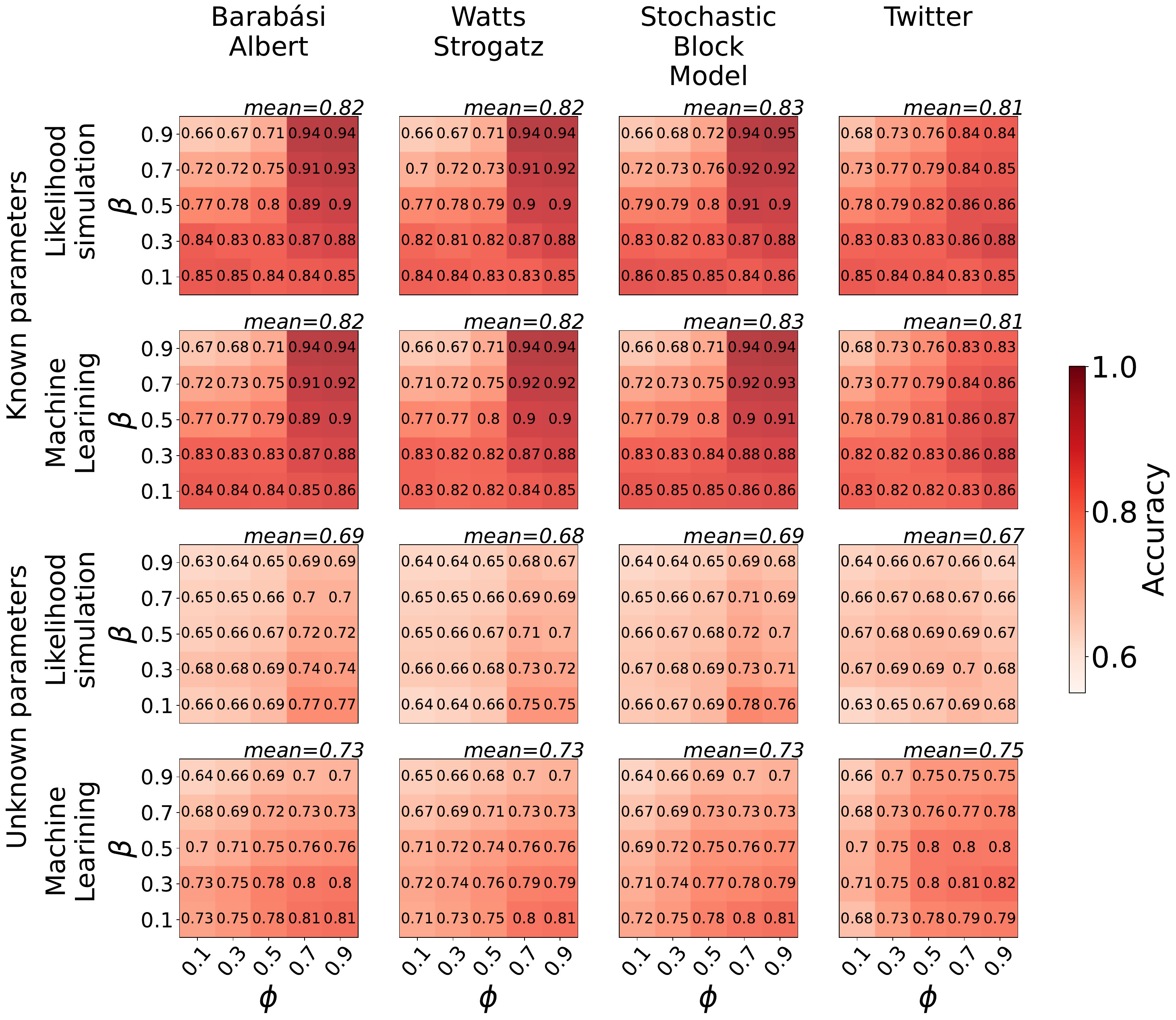}
\caption{Accuracies obtained by classifying the infection instances from Experiment 2 and 3 on different networks (x-axis) and with different methods (y-axis). The values of accuracies do not depend on the structure of the network.}
\label{fig_different_networks}
\end{figure}

To understand how network structure influences process distinguishability, we apply the classification methods on Experiment 2 and 3 on various networks (Figure \ref{fig_different_networks}). 
The values of accuracies remain consistent across the Barabási-Albert, Watts Strogatz and stochastic block model networks. 
However, we observe a decrease of 0.02 on the accuracy average considering a true Twitter network, but with the machine learning method with unknown parameters. 
Indeed, one of the most important feature of this method is the degree (Figure \ref{best_features}), which present larger variation with the Twitter network.


\section{Accuracies of the classification of spontaneous adoption on Experiment 3}

\begin{figure*}[h!]
\centering
\includegraphics[width=160mm]{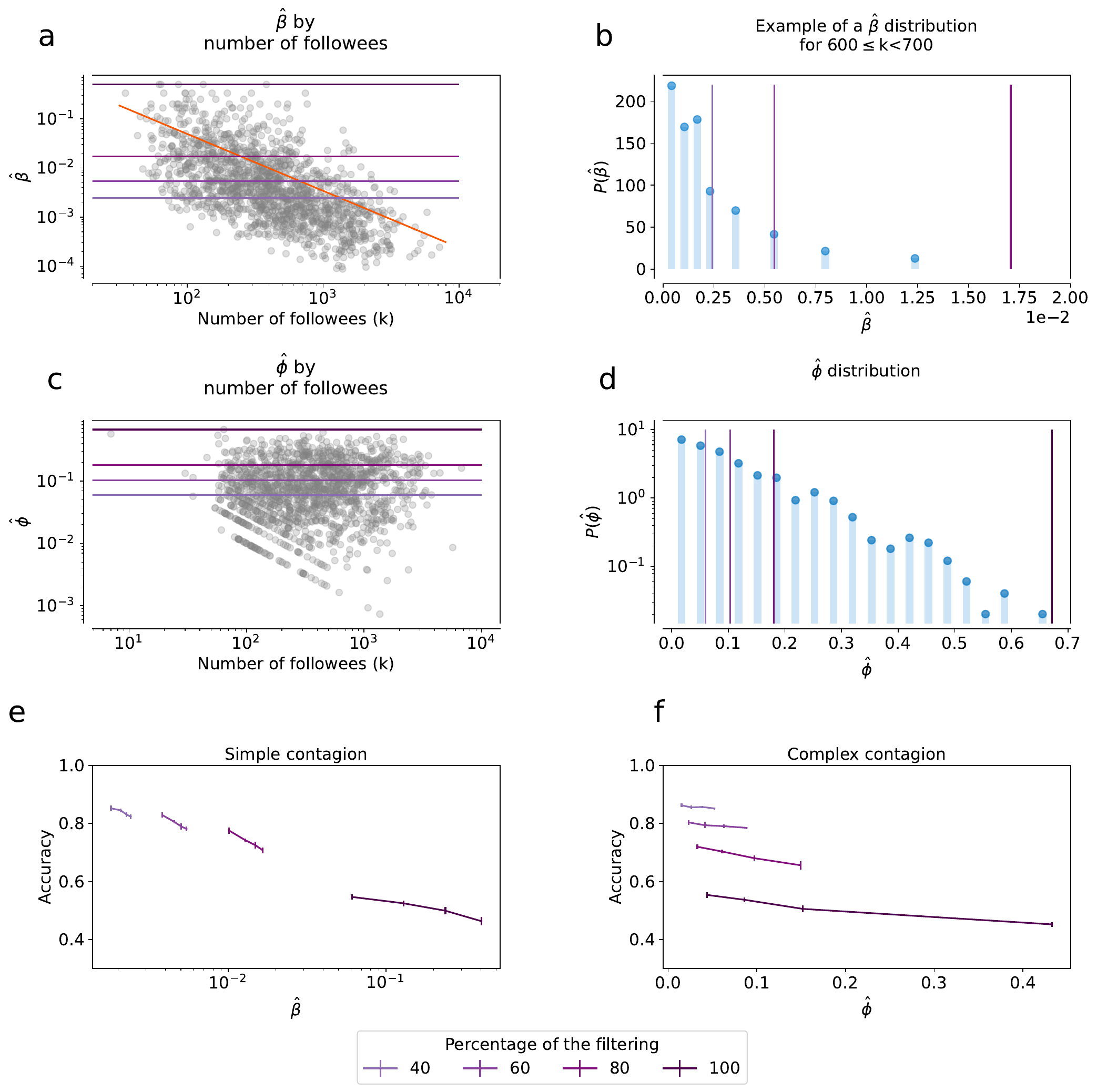}
\caption{Parametrization and accuracy of the classification of Experiment 4 constructed based on the \#GiletsJaunes Twitter dataset. The distributions of $\hat{\beta}$ and $\hat{\phi}$ respectively panels a-b and c-d are filtered keeping their 40\%, 60\%, 80\% or 100\% lower values. The accuracy values of the classification of the simple contagion (panel e) and the complex contagion (panel f) increase while the percentage of filtering increases.}
\label{ad_statistics_si}
\end{figure*}

We assess the classification accuracies of the simple, complex and spontaneous cases from Experiment 4 with the random forest algorithm, using different values of filtering on the values of $\hat{\beta}$ and $\hat{\phi}$ (40\%, 60\%, 80\% and 100\%). The accuracies of the classification of the simple and complex instances increase while the percentage of the filtering diminishes (Figure \ref{ad_statistics_si}), even though all the obtained values remain above the accuracy of the random classification (0.33). We choose to work with a filter of 80\% which presents accuracies above 0.65 while keeping most of the values of the distribution. The accuracies of the classification of the spontaneous instances (Table \ref{accuracy_st_exp3}) are consistently low, regardless the percentage of data filtering employed in Experiment 3. This is attributed to the inability to assess the rate of infection $r$.


    \begin{table}[]
    \begin{center}
\begin{tabular}{|l|l|l|l|l|}
\hline
Percentage of filtering & 40  & 60  & 80  & 100  \\ \hline
Accuracy of the St      & 0.07 & 0.12 & 0.23 & 0.50 \\ \hline
\end{tabular}
\caption{Accuracy of the classification of the spontaneous adoptions on the Experiment 4 with the random forest}
\label{accuracy_st_exp4}
\end{center}
\end{table}


\end{document}